\documentclass[reqno]{article}
\usepackage{amsmath,amssymb}
\usepackage{graphicx,color}
\usepackage[top=0.75in, bottom=0.75in, left=0.75in, right=0.75in, dvips]{geometry}
\usepackage{caption}
\usepackage{hyperref}
\usepackage[noblocks]{authblk}
\usepackage{epstopdf}
\begin{document}

\title{{\sf Can the Renormalization Group Improved Effective Potential be used to estimate the Higgs Mass in the Conformal Limit of the Standard Model? }}
\author[1]{F.A.~Chishtie}
\author[2,8]{T.~Hanif}
\author[1]{J.~Jia}
\author[3]{R.B.~Mann}
\author[1,4] {D.G.C.~McKeon}
\author[5,6,*]{T.N.~Sherry}
\author[7]{T.G.~Steele}
\affil[1] {Department of Applied Mathematics, The
University of Western Ontario, London, ON N6A 5B7, Canada}
\affil[2] {Department of Physics and Astronomy, The University of Western Ontario, London, ON N6A
5B7, Canada}
\affil[3] {Department of Physics, University of Waterloo, Waterloo, ON N2L
3G1, Canada}
\affil[4] {Department of Mathematics and
Computer Science, Algoma University, Sault St. Marie, ON N6A
2G4, Canada}
\affil[5] {School of Mathematics, Statistics and Applied Mathematics, NUI Galway, University Road, Galway, Ireland }
\affil[6] {School of Theoretical Physics, Dublin Institute for Advanced Studies, Burlington Rd., Dublin 4, Ireland }
\affil[7] {Department of Physics and Engineering Physics,
University of Saskatchewan, Saskatoon, SK S7N 5E2, Canada}
\affil[8] {Department of Theoretical Physics, University of Dhaka, Dhaka-1000, Bangladesh}

\maketitle

\maketitle

\let\oldthefootnote\thefootnote
\renewcommand{\thefootnote}{\fnsymbol{footnote}}
\footnotetext[1]{To whom correspondence should be addressed. Email: tom.sherry@nuigalway.ie}
\let\thefootnote\oldthefootnote

\begin{abstract}
We consider the effective potential $V$ in the standard model with a single Higgs doublet in the limit that the only mass scale $\mu$
present is radiatively generated.  Using a technique that has been shown to determine $V$ completely in terms of the renormalization
group (RG) functions when using the Coleman-Weinberg (CW) renormalization scheme, we first sum leading-log (LL) contributions to $V$
using the one loop RG functions, associated with five couplings (the top quark Yukawa coupling $x$, the quartic coupling of the
Higgs field $y$, the $SU(3)$ gauge coupling $z$, and the $SU(2) \times U(1)$ couplings $r$ and $s$).  We then employ the two loop
RG functions with the three couplings $x$, $y$, $z$ to sum the next-to-leading-log (NLL) contributions to $V$ and then the three to
five loop RG functions with one coupling $y$ to sum all the $N^2LL \ldots N^4LL$ contributions to $V$.  In order to compute these sums,
it is necessary to convert those RG functions that have been originally computed explicitly in the minimal subtraction (MS) scheme to
their form in the CW scheme. The Higgs mass can then be determined from the effective potential: the $LL$ result is
$m_{H}=219\;GeV/c^2$ decreases to $m_{H}=188\;GeV/c^2$ at $N^{2}LL$ order and $m_{H}=163\;GeV/c^2$ at $N^{4}LL$ order. No reasonable estimate of $m_H$ can be made at orders $V_{NLL}$ or $V_{N^3LL}$ since the method employed gives either negative or imaginary values for the quartic scalar coupling.  The fact that we get reasonable values for $m_H$  from the $LL$, $N^2LL$ and $N^4LL$ approximations
is taken to be an indication that this mechanism for spontaneous symmetry breaking  is in fact viable, though one in which there is slow
convergence towards the actual value of $m_H$.  The mass $163\;GeV/c^2$ is argued to be an upper bound on $m_H$.
\end{abstract}

\section*{Keywords}
Renormalization group; Effective potential; Standard model; Higgs mass; Coleman-Weinberg renormalization scheme; Radiative effects.

\section{Introduction}

The leading-logarithm (LL) contribution to the effective potential $V$ in the standard model in which there is a single scalar field and no mass scale in the classical limit, has been used to estimate the Higgs mass to be $m_H = 224\;GeV/c^2$ \cite{1}.
Subsequent investigations indicate that contributions beyond LL to $V$ do not destabilize this result \cite{2}.  In this paper we propose
to significantly improve the methods used in refs.~\cite{1, 2} and compute the resulting modification to the estimate of $m_H$. The value of $m_H$ obtained using these improvements is much more realistic.

Since these results were obtained, it has been established that when the CW renormalization scheme is used to compute $V$, all $N^pLL$ contributions
to $V$ can be computed using the ($p + 1$) loop RG functions when there is a single scalar field $\phi$ without a classical mass term for this
scalar in the action \cite{3}.

We first show how these techniques can be used to refine the approach of \cite{1, 2}.  In doing so, we overcome several shortcomings
of the original calculation.  First of all, the RG functions we use are those appropriate to the CW renormalization scheme, not the minimal subtraction
(MS) scheme.  This conversion from the MS scheme (in which the RG functions were originally computed) to the CW scheme was not carried out in
\cite{1, 2}.  Next, we show how the $N^pLL$ contributions to $V$ can be expressed exactly in terms of the ($p+1$) loop CW RG functions. This shows that once the $(p+1)$ loop CW RG functions are known, we have an exact expression for the $(p+1)$ loop contributions to $V$ without having to compute any Feynman diagrams and, in addition, we can sum all the $N^pLL$ contributions to $V$ coming from all orders in the loop expansion. In \cite{1,2} these
contributions were only given as a power series in the couplings $x$, $z$, $r$ and $s$.  Finally, we compute the counter-term that takes into
account all log-independent contributions to $V$ beyond the $N^pLL$ order in a more consistent way than was done in \cite{1, 2}; rather than
fixing this counter-term by the $LL$ calculation and then using this value at higher order, we determine the value of this counter-term at each
order separately thereby taking into account how the value of the coupling $y$ is adjusted.  It is the
methods of ref.~\cite{3} that allow us to fix all $\log$ independent contributions to $V$ in terms of the RG functions when using the CW scheme.  Our analytic approach  supplements  numerical techniques for investigating $V$ using the RG equation (see e.g., ref.~\cite{meissner}).

In the next section we review how $N^pLL$ contributions to $V$ can be computed in terms of the RG functions when the CW renormalization scheme is used, first
considering the case in which there is a single $O(N)$ scalar field with only a quartic self coupling and no classical mass term in the Lagrangian.
The only mass scale in such a theory is radiatively induced.  This is then extended so that the scalar couples to other fields (both vectors and
spinors).  The details of the solution at $NLL$ are presented in Appendix 1 along with an explanation of how the methodology can be extended to $N^2LL$ and higher-order.
Appendix 2 presents a method of computing terms in the derivative expansion of the one loop effective action.

We have employed the CW renormalization scheme, as in this scheme all logarithmic dependence on the external field comes through a single form of logarithm, $\ln\left(\phi^2/\mu^2\right)$. Having this single logarithm simplifies the ansatz we make for $V$ when there are multiple couplings (see 
eq.~(\ref{eq19}) below), making it possible to find $V$ in terms of the CW RG functions. If there are multiple couplings (say $x$ and $y$) then both $\ln\left(x\phi^2/\mu^2\right)$ and $\ln\left(y\phi^2/\mu^2\right)$ arise when using the MS renormalization scheme.  This complicates the ansatz one has for $V$, making
it no longer feasible to find $V$ in terms of the MS RG functions.  Furthermore, one must compute the radiative corrections dependent on $\phi$ to the kinetic term $(\partial_{\mu}\phi)^2$ in the effective Lagrangian when determining the radiatively generated Higgs mass $m_H$; this is unknown (and presumably non-trivial) in the MS scheme, whereas in the CW scheme it is defined to be equal to one at the value of $\phi$ that minimizes $V$ (see eq.~(\ref{eq18}) below).  For these reasons we use the CW scheme in our analysis.

    We also note that the inclusion of a quadratic mass term $m^{2}\phi^{2}$ for the $O(4)$ scalar field in the classical action results in multiple forms of the logarithm occurring in the ansatz for $V$ (see ref. \cite{10}) and also necessitates consideration of a ``cosmological term" (see ref.~\cite{kastening}). These factors considerably complicate employing the RG equation to find the $N^pLL$ contributions to $V$; we thus restrict ourselves to the classically conformal case $m=0$ as originally suggested in ref.~\cite{4}.

We then discuss the conversion of the RG functions from the MS scheme, in which they have been originally computed, to the CW scheme, which is necessary
to implement our procedure for computing the $N^pLL$ contribution to $V$.
We finally apply these results to the simplest version of the standard model in which there is a single scalar which is an $SU(2)$ doublet and which
has no mass at the classical level.  The resulting expression for the effective potential at $N^4LL$ order leads to an estimate of $163\;GeV/c^2$ for the
mass of the Higgs Boson.  We regard this  as an upper limit on the Higgs mass as lower order calculations lead to estimates that are
considerably higher than this.  In any case, the proposal \cite{4} that the Higgs mechanism is a consequence of radiative corrections to the effective potential in
the conformally invariant classical limit of the standard model is seen to be viable.

We note that the potential $V$ being considered here is the sum of all one particle irreducible (1PI) diagrams with external scalar fields whose momentum vanishes. This 1PI potential has been argued to be distinct from the ``effective potential", a quantity shown in ref.~\cite{26} to be convex and real. The relationship between the 1PI potential and the effective potential is discussed in refs.~\cite{27,28} and reviewed in refs.~\cite{29,30,31}. However,  resolution of the convexity problem continues to be debated in the literature (see refs.~\cite{32,32b,33}). The most recent examination of the convexity problem explores the distinctions between the Euclidean and Minkowskian formulations of the effective potential \cite{34}.

Although our work adopts the conventional approach of ascribing physical meaning to the 1PI potential \cite{27,28,29,30}, it is important to note that our Higgs mass predictions in the  standard model rely upon only the {\em local} properties of the 1PI potential near the minimum as extracted from the RG equation.   Since this minimum occurs at non-zero field values, the minimum corresponds to the qualitative non-perturbative form of a spontaneous symmetry breaking  effective potential \cite{callaway} and provides the lower bound  on the region where the effective potential and 1PI potential coincide \cite{27,28,30}.
 Therefore our analysis is not in conflict with Ref.~\cite{32}, which argues that the 1PI and effective potential must agree near the minimum and advocates the use of RG methods.

Finally we note that non-perturbative approaches are not isolated from the convexity problem. For example, the constraint effective potential \cite{35} in lattice approaches is non-convex at finite volumes \cite{36}, and lattice results are found to agree with the perturbative 1PI potential in appropriate regions of parameter space \cite{37}.  Functional flows of the exact renormalization group can be used to
calculate an effective {\em average} action \cite{wetterich,tetradis}  and convexity constrains the regulators  used in various  truncation schemes  used in these methods \cite{litim}.\footnote{The average effective action is calculated for scalar QED in Ref.~\cite{litim2}.}  Other alternatives to the effective potential include the Gaussian effective potential \cite{stephenson} which is well-suited to variational techniques.

\section{Summing Logarithms in the Effective Potential}

We begin by considering an $O(N)$ scalar field $\phi$ with a classical potential $V_{cl}$
\begin{equation}
V_{cl} = \lambda \phi^4 = \pi^2 y \phi^4
\label{eq1}
\end{equation}
where $\lambda$ is the usual scalar coupling constant but $y$ is more useful as it removes explicit factors of $\pi^2$ in RG functions.
The coupling $y$ is renormalized so that the effective potential $V$ satisfies the CW renormalization condition \cite{4}
\begin{equation}
\frac{d^4V(\phi)}{d\phi^4}\left|_{\phi = \mu} = 24 \pi^2 y \right.
\label{eq2}
\end{equation}
is satisfied.
Radiative corrections to the effective potential \cite{4,5,6,29} with this renormalization condition take the form
\begin{equation}
V(y,\phi , \mu) = \pi^2 \sum_{n=0}^\infty \sum_{m=0}^n y^{n+1} T_{nm}L^m\phi^4
\label{eq3}
\end{equation}
where $L = \ln\left(\frac{\phi^2}{\mu^2}\right)$.
In order that there be no net dependence on the renormalization scale parameter $\mu$, $V$ must satisfy
\begin{equation}
\mu \frac{dV}{d\mu} = 0 = \left(\mu \frac{\partial}{\partial\mu} + \beta (y) \frac{\partial}{\partial y} - \phi \gamma (y) \frac{\partial}{\partial\phi}\right) V
\label{eq4}
\end{equation}
where
\begin{equation}
\mu \frac{dy}{d\mu} = \beta (y)
= \sum_{n=2}^\infty b_ny^n
\label{eq5}
\end{equation}
and
\begin{equation}
\frac{\mu}{\phi}\frac{d\phi}{d\mu} = - \gamma(y)
= - \sum_{n=1}^\infty g_ny^n.
\label{eq6}
\end{equation}
The RG equation \eqref{eq4} and its solution for $V$ in eq.~\eqref{eq3} corresponds to  the situation where there is no quadratic term for the scalar field,  consistently maintaining the massless nature of the theory. In particular, extension to massive theories
is achieved   by including a mass term and anomalous mass dimension into the RG equation \eqref{eq4} (see Ref.~\cite{kastening} for an analysis of  a single-component massive  scalar theory).
It is therefore not necessary for us  to impose the $V^{\prime\prime}(\phi=0)=0$ renormalization condition used by Coleman \& Weinberg \cite{4} to eliminate quadratic divergences.  Furthermore, vacuum graphs do not generate divergences that are eliminated by renormalization of the cosmological term \cite{kastening}.
 As outlined below,  we also do not introduce quadratic counter-terms into the phenomenological analysis of $V$.

If now the $N^pLL$ contribution to $V$ in eq.~\eqref{eq3} is defined to be $V_{N^{p}LL} = \pi^2y^{p+1}S_p(yL)\phi^4$ where
\begin{equation}
S_n (yL) = \sum_{m=0}^\infty T_{n+m,m}(yL)^m
\label{eq7}
\end{equation}
so that
\begin{equation}
V = \pi^2 \sum_{n=0}^\infty y^{n+1} S_n(yL)\phi^4
\label{eq8}
\end{equation}
then eq.~\eqref{eq4} is satisfied at order $y^{n+2}$ provided $S_n(\xi)$ satisfies
\begin{equation}
\left[(-2+b_2\xi)\frac{d}{d\xi} + b_2 - 4g_1\right]S_0 = 0
\label{eq9}
\end{equation}
and
\begin{equation}
\left[(-2 + b_2\xi) \frac{d}{d\xi}  + (n+1) b_2 - 4g_1 \right]S_n + \sum_{m=0}^{n-1} \left[-2g_{n-m}+b_{n-m+2} \xi \frac{d}{d\xi} + (m+1)b_{n-m+2}-4g_{n-m+1} \right]S_m
= 0
\label{eq10}
\end{equation}
with the boundary condition
\begin{equation}
S_n(0) = T_{n0}.
\label{eq11}
\end{equation}
Thus $V$ can be determined by solving the coupled equations (\ref{eq9}, \ref{eq10}) provided the boundary values $T_{n0}$ are known.  These are fixed
by the CW condition of eq.~\eqref{eq2}; since $L = 0$ when $\phi = \mu$ eqs.~(\ref{eq2}, \ref{eq8}) together imply that
\begin{equation}
24y = \sum_{k=0}^\infty y^{k+1} \left[16y^4 S_k^{\prime\prime\prime\prime} (0) + 80y^3S_k^{\prime\prime\prime}(0) + 140 y^2S_k^{\prime\prime}(0) + 100 yS_k^\prime (0) + S_k(0)\right].
\label{eq12}
\end{equation}
Since $g_1 = 0$, together \eqref{eq9} and \eqref{eq12} lead to
\begin{gather}
T_{00} = 1
\label{eq13}
\\
S_0(\xi) = \frac{1}{w}
\label{eq14}
\end{gather}
where $w = 1 - \frac{1}{2}b_2\xi$.  Eq.~\eqref{eq12} then gives
\begin{equation}
T_{10} = -\frac{25}{12} b_2
\label{eq15}
\end{equation}
so that eq.~\eqref{eq10} be solved when $n = 1$
\begin{equation}
\begin{split}
S_1(\xi) &= \frac{4g_2}{b_2w} - \frac{4g_2 + \frac{25}{12} b_2^2}{b_2w^2} - \frac{b_3}{b_2w^2} \ln |w|\\
&= \frac{1}{4w} + \left( \frac{1}{4} \ln |w| - \frac{51}{4}\right) \frac{1}{w^2}\;\;(\mathrm{for}\;N = 4).
\end{split}
\label{eq16}
\end{equation}
This process can be continued indefinitely; $S_p(\xi)$ can be determined in terms of $b_2 \ldots b_{p+2}$, $g_2 \ldots g_{p+1}$ where
these RG function coefficients are those appropriate to the CW scheme.

If in addition to $y$ there are other couplings $g_i (i = 1 \ldots N)$ (Yukawa, gauge etc.) in the theory then the CW renormalization condition \eqref{eq2} must be
supplemented by additional conditions.  For example, in massless scalar electrodynamics in which a complex scalar $\phi$ is coupled to a $U(1)$ gauge field $A_\mu$ with
coupling $e$, then the effective action takes the form \cite{4}
\begin{equation}
\Gamma = \int d^4x \left[ -V(\phi) + \frac{1}{2} Z(\phi)\left|(\partial_\mu - ieA_\mu)\phi\right|^2 -  \frac{1}{4} H(\phi)(\partial_\mu A_\nu - \partial_\nu A_\mu)^2 + \ldots\right].
\label{eq17}
\end{equation}
Infinities arise when computing $V$, $Z$ and $H$ and so in addition to \eqref{eq2} one requires renormalization conditions which we take to be
\begin{equation}
H(\phi = \mu) = 1 = Z(\phi = \mu).
\label{eq18}
\end{equation}
Application of the RG equation to determine higher order corrections to $Z(\phi)$ is discussed in ref.~\cite{21}.

Suppose that $x$ and $y$ are the only two couplings. (It is easy to extend our considerations to include more than two.)  The expansion
of eq.~\eqref{eq3} now generalizes to
\begin{equation}
V = \pi^2 \sum_{n=1}^\infty \sum_{r=0}^{n+k} \sum_{k=0}^\infty T_{n+k-r,r,k}\,y^{n+k-r}x^rL^k
\label{eq19}
\end{equation}
and $V$ satisfies the RG equation
\begin{equation}
\left( \mu \frac{\partial}{\partial\mu} + \beta^x \frac{\partial}{\partial x} + \beta^y \frac{\partial}{\partial y} - \phi\gamma \frac{\partial}
{\partial \phi} \right) V=0.
\label{eq20}
\end{equation}
The RG functions are
\begin{align}
\beta^x &= \mu \frac{dx}{d\mu} = \sum_{n=2}^\infty \beta_n^x = \sum_{n=2}^\infty \sum_{r=0}^n b_{n-r,r}^{\;x} \,x^ry^{n-r}
\label{eq21}
\\
\beta^y &= \mu \frac{dy}{d\mu} = \sum_{n=2}^\infty \beta_n^y = \sum_{n=2}^\infty \sum_{r=0}^n b_{n-r,r}^y \,x^ry^{n-r}
\label{eq22}
\\
\gamma &= -\frac{\mu}{\phi} \frac{d\phi}{d\mu} = \sum_{n=1}^\infty \gamma_n = \sum_{n=1}^\infty \sum_{r=0}^n g_{n-r,r}\,x^ry^{n-r}.
\label{eq23}
\end{align}
The $N^pLL$ contribution to $V$ is now given by
\begin{equation}
V_{N^pLL} = \pi^2 \sum_{k=0}^\infty p_{k+p+1}^{\,k} L^k\phi^4
\label{eq24}
\end{equation}
where
\begin{equation}
p_n^k(x,y) = \sum_{r=0}^n T_{n-r,r,k}y^{n-r}x^r\quad (n \geq k + 1)
\label{eq25}
\end{equation}
so that
\begin{equation}
V = \sum_{p=0}^\infty V_{N^pLL}.
\label{eq26}
\end{equation}

The CW condition of eq.~\eqref{eq2} now shows that for all $n$
\begin{equation}
24y \delta_{n0} = 24p_n^0 + 100p_n^1 + 280p_n^2 + 480 p_n^3 +384 p_n^4.
\label{eq27}
\end{equation}
Furthermore, the RG equation \eqref{eq20} leads to
\begin{equation}
\sum_{n=1}^\infty \sum_{k=0}^{n-1} \left[-2kp_n^k L^{k-1} + \sum_{m=2}^\infty \left(\beta_m^x \frac{\partial}{\partial x} + \beta_m^y \frac{\partial}{\partial y}\right)
p_n^k L^k - \sum_{m=1}^\infty \left(4\gamma_m p_n^k L^k + 2k\gamma_mp_n^k L^{k-1}\right)\right] =0.
\label{eq28}
\end{equation}
Together, (\ref{eq27}, \ref{eq28}) fix $V$ in terms of the CW RG functions.

 We employ a novel way of treating the sums in eq. (\ref{eq24}), which involves using the method of characteristics \cite{3}. Beginning with the definition
\begin{equation}
w_n^k (\overline{x}(t), \overline{y}(t), t) = p_n^k (\overline{x}(t), \overline{y}(t))\exp \left[-4 \int_0^t \gamma_1(\overline{x}(\tau),\overline{y}(\tau))d\tau\right]
\label{eq29}
\end{equation}
where
\begin{align}
\frac{d\overline{x}(t)}{dt} &= \beta_2^x (\overline{x}(t),\overline{y}(t))
\label{eq30}
\\
\frac{d\overline{y}(t)}{dt} &= \beta_2^y (\overline{x}(t),\overline{y}(t))
\label{eq31}
\end{align}
with $\overline{x}(0) = x$, $\overline{y}(0) = y$ we find that
\begin{equation}
\frac{d}{dt} w_n^k (\overline{x}, \overline{y}, t) = \left(\beta_2^x (\overline{x}, \overline{y}) \frac{\partial}{\partial\overline{x}}
+ \beta_2^y(\overline{x}, \overline{y})\frac{\partial}{\partial \overline{y}} - 4 \gamma_1(\overline{x}, \overline{y})\right) w_n^k(\overline{x}, \overline{y}, t).
\label{eq32}
\end{equation}
Eq.~\eqref{eq28} is satisfied to order $n - 1$ in $L$ and $n + 1$ in the couplings $x$ and $y$ provided
\begin{equation} p_{n+1}^n = \frac{1}{2n} \left(\beta_2^x \frac{\partial}{\partial x} + \beta_2^y \frac{\partial}{\partial y} - 4\gamma_1\right)p_n^{n-1}
\label{eq33}
\end{equation}
so that by eqs.~(\ref{eq29}, \ref{eq32}, \ref{eq33})
\begin{equation}
w_{n+1}^n (\overline{x}, \overline{y}, t) = \frac{1}{2n} \frac{d}{dt} w_n^{n-1} (\overline{x},
\overline{y}, t).
\label{eq34}
\end{equation}
If now
\begin{equation}
\overline{V}_{N^pLL}(\overline{x}, \overline{y}, t) = \pi^2 \sum_{k=0}^\infty w_{k+p+1}^k (\overline{x}, \overline{y}, t)L^k\phi^4
\label{eq35}
\end{equation}
so that if $t = 0$
\begin{equation}
\overline{V}_{N^pLL}(x, y, 0) = V_{N^pLL}
\label{eq36}
\end{equation}
then by \eqref{eq29}
\begin{equation}
\overline{V}_{LL}(\overline{x}(t), \overline{y}(t), t)
= \pi^2 \sum_{n=0}^\infty \frac{L^n}{2^nn!} \frac{d^n}{dt^n}
w_1^0 (\overline{x}(t), \overline{y}(t), t)\phi^4
= \pi^2 w_1^0 (\overline{x}(t + \frac{L}{2}), \overline{y}(t+\frac{L}{2}), \frac{L}{2})\phi^4
\label{eq37}
\end{equation}
and hence by \eqref{eq36} we finally have a closed form expression for $V_{LL}$.
\begin{equation}
V_{LL} = \pi^2 w_1^0 (\overline{x}(\frac{L}{2}), \overline{y}(\frac{L}{2}), \frac{L}{2})\phi^4.
\label{eq38}
\end{equation}

The detailed computation of $V_{NLL}$ presented in Appendix 1 gives eq.~\eqref{eq64}
\begin{equation}
\begin{split}
V_{NLL}&= \pi^2\phi^4 \exp \left[ -4 \int_0^{L/2} d\tau \gamma_1 (\overline{x}^i (\tau)) \right]\Biggl\{ p_2^0 \left(\overline{x}^i\left(\frac{L}{2}\right)\right)  \Biggr.
\\
&\qquad + \int_0^{L/2} d\tau\left[ \left(-\gamma_1(\overline{x}^i(\tau))\beta_2^{x^{i}}(\overline{x}^i(\tau)) + \beta_3^{x^{i}} (\overline{x}^i(\tau))\right)\mathbf{U}_{ij}(0,\tau)\right].\left[ \mathbf{U}_{jk}\left(\frac{L}{2} ,0\right)\frac{\partial}{\partial\overline{x}^k(\frac{L}{2})} p_1^0 \left(\overline{x}^i\left(\frac{L}{2}\right)\right)\right]
\\
&\qquad\qquad\qquad \Biggl. + 4 \int_0^{L/2} d\tau \left[\gamma_1^2 (\overline{x}^i(\tau)) - \gamma_2(\overline{x}^i(\tau))\right] p_1^0\left(\overline{x}^i\left(\frac{L}{2}\right)\right)\Biggr\} ,
\end{split}
\label{neweq64}
\end{equation}
where by eqs.~(\ref{eq27}, \ref{eq33}, \ref{eq39})
\begin{equation}
p_1^0=y~,~p_2^1=\frac{1}{2}\beta_2^y-2\gamma_1 y~,~p_2^0=-\frac{25}{6}p_2^1,
\label{neweq40}
\end{equation}
and by eqs.~(\ref{eq51}--\ref{eq55})
\begin{gather}
\frac{d}{dt}\mathbf{U} (t,0) = \mathbf{U} (t,0)  \mathbf{M}
\\
\mathbf{U}^{-1}(t,0) = \mathbf{U}(0,t) = 1 + \sum_{n=1}^\infty (-1)^n \int_0^t d\tau_1 \ldots \int_0^{\tau_{n-1}} d\tau_n
\left[ \  \mathbf{M}(\tau_1) \ldots  \mathbf{M}(\tau_n) \  \right]
\end{gather}
and
\begin{equation}
\mathbf{M}_{ij}=\frac{\partial\beta_2^{x^j}}{\partial\bar x^i }.
\end{equation}
The  techniques used to find $V_{NLL}$ in eq.~\eqref{neweq64} can be extended to obtain $V_{N^{2}LL}$. However, since the three
loop RG functions needed for this extension have not been computed for the standard model, we will not pursue this calculation further.

We now will discuss how the CW RG functions can be found if the MS RG functions are known.

\section{Finding the Coleman-Weinberg Renormalization Group Functions}

The RG functions have been computed using dimensional regularization and minimal subtraction to five loop order in an $O(N)$ scalar theory \cite{8} and to two loop
order in the standard model \cite{9}.  We will now examine how from these known results one can find the RG functions in the CW renormalization scheme.

First, we quote the MS values of the $O(N)$ scalar model of eq.~\eqref{eq1} to five loop order \cite{8}
\begin{equation}
\begin{split}
\hspace{-.2cm}\tilde{\beta}(y) =& \frac{N+8}{2} y^2 -\frac{3}{4} (3N + 14)y^3 + \frac{1}{64} \left[33N^2 + 922N + 2960 + 96 (5N + 22) \zeta(3)\right]y^4
\\
&  - \frac{4}{3} \left(\frac{3}{2}\right)^5 \frac{y^5}{7776}
\biggl[-5N^3 + 6320N^2  + 80456N + 196648 + 96 \left(63N^2 + 764N + 2332\right)\zeta (3)
\\
& \qquad\qquad  \Biggl.-288 (5N + 22)(N+8)\zeta (4) + 1920 \left(2N^2 + 55N + 186\right)\zeta(5)]\Biggr]
\\
&   + \frac{4}{3} \left(\frac{3}{2}\right)^6 \frac{y^6}{124416} \Biggl[ 13N^4 + 12578N^3 + 808496N^2 + 6646336N + 13177344  \Biggr.
\\
& \quad + 16 \left(-9N^4 + 1248N^3 + 67640N^2 + 552280N +  1314336\right)\zeta(3)
+ 768\left(-6N^3 - 59N^2 + 446N + 3264\right)\zeta^2(3)
\\
&\quad- 288\left(63N^3 + 1388N^2 + 9532N + 21120\right)\zeta (4)
+ 256 \left(305N^3 + 7466N^2 + 66986N + 165084\right)\zeta (5)
\\
&\quad \Biggl.- 9600 (N+8) \left(2N^2 + 55N + 186\right)\zeta (6)
 + 112896 \left(14N^2 + 189N + 526\right) \zeta (7)\Biggr] + \mathcal{O}(y^7)
\end{split}
\label{eq68}
\end{equation}
and
\begin{equation}
\begin{split}
\hspace{-.5cm}\tilde{\gamma}(y) &= \frac{N+2}{16} y^2 - \frac{(N+2)(N+8)}{128} y^3 + \left(\frac{3}{2}\right)^4 \frac{y^4}{5184} (N+2)\left[5\left(-N^2
+ 18N + 100\right)\right]
\\
& - \left(\frac{3}{2}\right)^5 \frac{y^5}{186624} (N+2) \Biggl[39 N^3 + 296N^2
+ 22752N + 77056 - 48(N^3 - 6N^2 + 64N + 184)\zeta(3)\Biggr.
\\
&\qquad\qquad\qquad\qquad  \Biggl.+ 1152(5N + 22) \zeta(4)\Biggr] + \mathcal{O}(y^6).
\end{split}
\label{eq69}
\end{equation}

We next provide the two loop RG functions in the standard model in which there is a single scalar doublet with no mass term for this
field in the classical action.  The quartic scalar coupling $y$ appears in eq.~\eqref{eq1}; the other couplings are the top quark Yukawa coupling
\begin{equation}
x = \frac{g_t^2}{4\pi^2}
\label{eq70}
\end{equation}
the $SU(3)$ coupling
\begin{equation}
z = \frac{g_3^2}{4\pi^2}
\label{eq71}
\end{equation}
and the $SU(2) \times U(1)$ couplings
\begin{gather}
r = \frac{g_2^2}{4\pi^2}
\label{eq72}
\\
s = \frac{g_1^2}{4\pi^2}\;.
\label{eq73}
\end{gather}
To two loop order the RG functions in this simplest version of the standard model \cite{9} in the MS renormalization scheme are
{\allowdisplaybreaks
\begin{gather}
\begin{split}
\tilde{\beta}^x =& \tilde{\mu} \frac{dx}{d\tilde{\mu}}=\left[\frac{9}{4} x^2 - 4xz - \frac{9}{8} xr - \frac{17}{24} xs\right]
+ \Biggl[ -\frac{3}{2} x^3 + \frac{131}{128} x^2s + \frac{225}{128} x^2r + \frac{9}{2}x^2z - \frac{3}{2} x^2y \Biggr.  \\
&\quad\qquad\qquad  \Biggl.+ \frac{1187}{1728} xs^2 - \frac{3}{32} xrs + \frac{19}{72}xsz - \frac{23}{32} xr^2
 +\frac{9}{8} xrz - \frac{27}{2} xz^2 + \frac{3}{4}xy^2 \Biggr]  + \ldots
\end{split}
\label{eq74}
\\
\begin{split}
\tilde{\beta}^y =& \tilde{\mu} \frac{dy}{d\tilde{\mu}} = \left[6y^2 + 3xy - \frac{3}{2} x^2 -  \frac{9}{4} yr - \frac{3}{4} ys + \frac{3}{32} s^2
+ \frac{3}{16} rs + \frac{9}{32} r^2\right]\\
& + \Biggl[ -\frac{39}{2} y^3 - 9xy^2 + \frac{27}{4} y^2r + \frac{9}{4} y^2s - \frac{3}{16}x^2y + 5xyz  + \frac{45}{32} xyr + \frac{85}{96} xys - \frac{73}{128}yr^2 + \frac{39}{64} yrs + \frac{629}{384} ys^2 \Biggr.
\\
&\quad \Biggl. + \frac{15}{8} x^3 - 2x^2z - \frac{1}{6} x^2s - \frac{9}{64} xr^2 + \frac{21}{32}xrs - \frac{19}{64} xs^2
+\frac{305}{256} r^3 - \frac{289}{768} r^2s - \frac{559}{768}rs^2 - \frac{379}{768} s^3 \Biggr]  + \ldots
\end{split}
\label{eq75}
\\
\tilde{\beta}^z = \tilde{\mu} \frac{dz}{d\tilde{\mu}} = \left[ - \frac{7}{2} z^2\right]+
\left[\frac{11}{48} sz^2 + \frac{9}{16} rz^2 - \frac{13}{4}z^3 - \frac{xz^2}{4}\right] + \ldots
\label{eq76}
\\
\tilde{\beta}^r = \tilde{\mu} \frac{dr}{d\tilde{\mu}} = \left[ - \frac{19}{12} r^2\right]
+ \left[\frac{3}{16} r^2s + \frac{35}{48} r^3 + \frac{3}{2}r^3z - \frac{3}{16} xr^2\right] + \ldots
\label{eq77}
\\
\tilde{\beta}^s = \tilde{\mu} \frac{ds}{d\tilde{\mu}} =\left[  \frac{41}{12} s^2\right] +
\left[\frac{199}{144} s^3 + \frac{9}{16} rs^2 + \frac{11}{6}zs^2 - \frac{17}{48} xs^2\right] + \ldots
\label{eq78}
\end{gather}
}
and
\begin{equation}
\tilde{\gamma} = -\frac{\tilde{\mu}}{\phi} \frac{d\phi}{d\tilde{\mu}} = \left[\frac{3}{4}x - \frac{9}{16}r - \frac{3}{16} s\right] +
\left[\frac{3}{8} y^2 -  \frac{27}{64} x^2 + \frac{5}{4} xz
+\frac{45}{128} xr +  \frac{85}{384} xs - \frac{271}{512} r^2 + \frac{9}{256}rs + \frac{41}{1536} s^2 \right]
+ \ldots
\label{eq79}
\end{equation}

In the case of there being only an $O(N)$ scalar field $\phi$, we follow the procedure outlined in refs.~\cite{3,22} to convert from the RG functions of
eqs.~(\ref{eq68}, \ref{eq69}) to those appropriate to the CW scheme.  In the MS scheme, the computation results in an expansion of $V$ that is similar to that of eq.~\eqref{eq3},
\begin{equation}
V = \pi^2 \sum_{n=0}^\infty \sum_{m=0}^n y^{n+1} \tilde{T}_{nm} \tilde{L}^m \phi^4
\label{eq80}
\end{equation}
where now $\tilde{L} = \ln \left(\frac{y\phi^2}{\tilde{\mu}^2}\right)$.  If the RG scale $\tilde{\mu}$ in the MS scheme is rescaled
\begin{equation}
\tilde{\mu} = y^{1/2} \mu
\label{eq81}
\end{equation}
where $\mu$ is the RG scale in the CW scheme, then the form of the expansion of eq.~\eqref{eq80} becomes that of eq.~\eqref{eq3}.  Finite renormalizations of the form
\begin{gather}
y  \rightarrow y(1 + a_1 y + a_2y^2 + \ldots)
\label{eq82}
\\
\phi  \rightarrow \phi (1 + b_1y + b_2y^2 + \ldots)
\label{eq83}
\end{gather}
may then be required to adjust the coefficients $\tilde{T}_{n0}$ in eq.~\eqref{eq80} so that the CW RG condition of eq.~\eqref{eq2} is satisfied, but this
can be done without altering $\tilde{T}_{nm} (m > 0)$ and hence the terms in $V$ that fix the RG functions are not changed \cite{11}.

With the rescaling of eq.~\eqref{eq81}
\begin{equation}
\beta(y) = \mu \frac{\partial y}{\partial \mu} = (\tilde{\mu} y^{-1/2})\left( \frac{\partial(y^{1/2}\mu)}{\partial \mu}\right)
\frac{\partial y}{\partial \tilde{\mu}}
= \tilde{\beta}(y)/\left(1 - \tilde{\beta}(y)/(2y)\right)
\label{eq84}
\end{equation}
and similarly
\begin{equation}
\gamma(y) = \tilde{\gamma}(y)/ \left(1-\tilde{\beta}(y)/(2y)\right).
\label{eq85}
\end{equation}
Eqs.~(\ref{eq84}, \ref{eq85}) allow one to pass from the MS RG functions of eqs.~(\ref{eq68}, \ref{eq69}) to the CW RG functions.

It is somewhat more complicated to convert the RG functions of eqs.~(\ref{eq69}--\ref{eq74}) to the CW scheme since more than one type of logarithm arises when $V$
is computed using the MS renormalization scheme.  A computation of $V$ in the CW scheme would allow one to infer the CW RG functions, but to obtain in
this way the RG functions to order $n$, one must compute $V$ to order $(n + 1)$ \cite{11}.  Since $V$ in the standard model has only been computed to second
order \cite{12} one cannot determine the CW RG functions to two loop order from $V$ directly; other contributions to the effective action must be considered.

Suppose the couplings in a theory are $g_i$ (with $g_i = (x,y,z,r,s)$ in the standard model) and that there is one scalar field $\phi$.  When
computing $V$ using MS, logarithms of the form $\tilde{L}_i = \ln\left(g_i\phi^2/\tilde{\mu}^2\right)$ arise.  At one loop order in MS, only these types of logarithms occur; beyond one loop order other more complicated logarithms arise \cite{12} but do not affect our discussion of how the MS and CW RG functions are related at two loop order. As in refs.~\cite{3,10} we associate
a separate renormalization scale $\kappa_i$ with each of these logarithms so that now
\begin{equation}
\tilde{L}_i = \ln\left(\frac{g_i\phi^2}{\kappa_i^2}\right).
\label{eq86}
\end{equation}
A rescaling similar to that of eq.~\eqref{eq81}
\begin{equation}
\kappa_i = g_i^{1/2}\mu
\label{eq87}
\end{equation}
leads to
\begin{gather}
\beta^{g_{i}} = \mu \frac{\partial g_i}{\partial\mu} = \sum_j \tilde{\beta}_j^{g_{i}} \left( 1 + \frac{\beta^{g_{j}}}{2g^j}\right)
\label{eq88}
\\
\gamma = -\frac{\phi}{\mu} \frac{\partial \phi}{\partial\mu} = \sum_j \tilde{\gamma}_j \left( 1 + \frac{\beta^{g_{j}}}{2g^j}\right)
\label{eq89}
\end{gather}
where
\begin{gather}
\tilde{\beta}_j^{g_{i}} = \kappa_j \frac{\partial g_i}{\partial\kappa_j}
\label{eq90}
\\
\tilde{\gamma}_j = - \frac{\kappa_j}{\phi}\frac{\partial\phi}{\partial\kappa_j}.
\label{eq91}
\end{gather}
Again, $\mu$ is the CW mass parameter.  We also see that
\begin{align}
\tilde{\beta}^{g_{i}} &= \sum_j \tilde{\beta}_j^{g_{i}}
\label{eq92}
\\
\tilde{\gamma} &= \sum_j\tilde{\gamma}_j
\label{eq93}
\end{align}
where $\tilde{\beta}^{g_{i}}$ and $\tilde{\gamma}$ are the MS RG functions.

We now will use eqs.~(\ref{eq88}, \ref{eq89}) to find the CW RG functions to two loop order in the standard model, restricting ourselves to the limiting case in which
only the three dominant couplings $g_1 = x$, $g_2 = y$ and $g_3 = z$ are considered.  If we use Roman numeral subscripts with the RG functions to denote the
number of coupling constants present in a perturbative expansion (e.g., $\tilde{\beta}_{1II}^x$ is the term in the expansion of the $\beta$ function for $x$
in the MS scheme associated with the mass scale $\kappa_1$ that has two powers of the coupling), then by eqs.~(\ref{eq88}--\ref{eq91}) we see that
\begin{align}
\beta_{II}^{g_{i}} = \tilde{\beta}_{II}^{g_{i}}
\label{eq94}
\\
\gamma_I = \tilde{\gamma}_I ;
\label{eq95}
\end{align}
that is at lowest order the RG functions in the CW and MS schemes are the same.  It also follows that
\begin{align}
\beta_{III}^{g_{i}} = \tilde{\beta}_{III}^{g_{i}} + \sum_j \frac{\tilde{\beta}_{j\,II}^{g_{i}}\beta_{II}^{g_{j}}}{2g_j}
\label{eq96}
\\
\gamma_{II} = \tilde{\gamma}_{II} + \sum_j \frac{\tilde{\gamma}_{j\,I}\beta_{II}^{g_{j}}}{2g_j} .
\label{eq97}
\end{align}
Eqs.~(\ref{eq96}, \ref{eq97}) show that apart from standard RG functions, only the one loop multi-scale RG quantities $\tilde{\beta}_{j\,II}^{g_{i}}$ and $\tilde{\gamma}_{j\,I}$ are needed to obtain the two loop CW RG functions
$\beta_{III}^{g_{i}}$ and $\gamma_{II}$.

To find $\tilde{\gamma}_{j\,I}$ we note that the one loop scalar self energy in the standard model (with no classical mass term for the
scalar and just the couplings $x$, $y$ and $z$) only has a contribution coming from the top quark loop.  Consequently the term $Z(\phi)(\partial_\mu\phi)^2$ in the
effective action only receives a logarithmic contribution of the form $\ln\left(x\phi^2/\tilde{\mu}^2\right)$ and so we see that
\begin{gather}
\tilde{\gamma}_{1\,I} = \tilde{\gamma}_I
\label{eq98}
\\
\tilde{\gamma}_{2\,I} = \tilde{\gamma}_{3\,I} = 0.
\label{eq99}
\end{gather}

To obtain $\tilde{\beta}_{1\,I}^y$, $\tilde{\beta}_{2\,II}^y$ and $\tilde{\beta}_{3\,II}^y$, we note that at leading log one loop order in the model
we are considering \cite{9}, $V$ is given in the MS scheme by
\begin{equation}
V = \pi^2 \left[y +\left(3y^2 \ln \frac{y\phi^2}{\tilde{\mu}^2} - \frac{3}{4} x^2 \ln \frac{x\phi^2}{\tilde{\mu}^2}\right)\right]\phi^4.
\label{eq100}
\end{equation}
If the RG equation of eq.~\eqref{eq20} is to be satisfied for each of the three mass scales $\kappa_j$ introduced in eq.~\eqref{eq86}, we find that consistency with eqs.~(\ref{eq98}, \ref{eq99}) occurs if
\begin{align}
\tilde{\beta}_{1\,II}^y &= -\frac{3}{2} x^2 + 3xy
\label{eq101}
\\
\tilde{\beta}_{2\,II}^y &= 6y^2
\label{eq102}
\end{align}
and
\begin{equation}
\tilde{\beta}_{3\,II}^y = 0.
\label{eq103}
\end{equation}

Determining $\tilde{\beta}_{1\,II}^z$, $\tilde{\beta}_{2\,II}^z$ and $\tilde{\beta}_{3\,II}^z$ is most easily done by considering the one
loop contribution to the term $-\frac{1}{4} H(\phi)F^2$ in the effective action where $F_{\mu\nu}^a$ is the $SU(3)$ field strength.  As only a quark
loop can contribute at one loop order to $H(\phi)$, then the only logarithmic contribution to $H(\phi)$ at one loop order is
$\ln \left(x\phi^2/\tilde{\mu}^2\right)$ in the MS scheme.  However, $H(\phi)$ dictates the function $\tilde{\beta}^z$ on account of gauge invariance \cite{13} and so
\begin{equation}
\tilde{\beta}_{1\,II}^z = \tilde{\beta}_{II}^z
\label{eq104}
\end{equation}
and
\begin{equation}
\tilde{\beta}_{2\,II}^z  = \tilde{\beta}_{3\,II}^z = 0.
\label{eq105}
\end{equation}

For $\tilde{\beta}_{1\,II}^x$, $\tilde{\beta}_{2\,II}^x$ and $\tilde{\beta}_{3\,II}^x$ we note that the scalar-quark-quark vertex only receives a logarithmic
contribution at one loop order of the form $\ln \left(x\phi^2/\tilde{\mu}^2\right)$ and hence
\begin{gather}
\tilde{\beta}_{1\,II}^x = \tilde{\beta}_{II}^x
\label{eq106}
\\
\tilde{\beta}_{2\,II}^x =  \tilde{\beta}_{3\,II}^x = 0.
\label{eq107}
\end{gather}

Together, eqs.~(\ref{eq98}--\ref{eq107}) result in eqs.~(\ref{eq96}, \ref{eq97}) yielding to two loop order in the CW scheme
\begin{gather}
\begin{split}
\beta^x =& \left[\frac{9}{4} x^2 -4xz\right] + \left[- \frac{3}{2} x^3 + \frac{9}{2} x^2z - \frac{3}{2} x^2y - \frac{27}{2} xz^2 + \frac{3}{4} xy^2\right]  + \frac{1}{2x}\left[\frac{9}{4} x^2 - 4xz\right]^2 + \ldots
\\
=& \left[\frac{9}{4} x^2 - 4xz\right]  + \left[ \frac{33}{32} x^3 - \frac{9}{2} x^2z - \frac{3}{2} x^2y +  \frac{3}{4} xy^2 - \frac{11}{2} xz^2 \right] + \ldots \\
\end{split}
\label{eq108}
\\
\begin{split}
\beta^y =& \left[6y^2 + 3xy - \frac{3}{2} x^2\right] + \left[-\frac{39}{2} y^3 - 9xy^2 - \frac{3}{16} x^2y + 5xyz  + \frac{15}{8} x^3 - 2 x^2z\right]
\\
& \quad + \frac{1}{2x} \left[- \frac{3}{2} x^2 + 3xy\right]\left[ \frac{9}{4} x^2 - 4xz\right]
+ \frac{1}{2y} \left[6y^2\right]\left[6y^2 + 3xy - \frac{3}{2} x^2\right] + \ldots
\\
=& \left[6y^2 + 3xy - \frac{3}{2} x^2\right] + \left[-\frac{3}{2} y^3 +  \frac{3}{16} x^3 + x^2z - xyz - \frac{21}{16} x^2y\right] + \ldots
\end{split}
\label{eq109}
\end{gather}
(which is the same result as is obtained from eq.~\eqref{eq84} if $x = z = 0$)
\begin{equation}
\begin{split}
\beta^z =& \left[-\frac{7}{2} z^2\right] + \left[-\frac{13}{4} z^3 -  \frac{1}{4} xz^2\right]
+  \frac{1}{2x} \left[- \frac{7}{2} z^2\right]\left[ \frac{9}{4} x^2 - 4xz\right] + \ldots
\\
= &\left[-\frac{7}{2} z^2\right] + \left[\frac{15}{4} z^3 -  \frac{67}{16} xz^2\right] + \ldots
\end{split}
\label{eq110}
\end{equation}
and
\begin{equation}
\begin{split}
\gamma &= \left[\frac{3}{4} x\right] + \left[-\frac{27}{64} x^2 +  \frac{3}{8} y^2 + \frac{5}{4} xz\right] +
\frac{1}{2x} \left[\frac{3}{4} x\right]\left[ \frac{9}{4} x^2 - 4xz\right] + \ldots
\\
&= \left[\frac{3}{4} x\right] + \left[\frac{27}{64} x^2 +  \frac{3}{8} y^2 - \frac{xz}{4} \right] + \ldots
\end{split}
\label{eq111}
\end{equation}
(Exact solutions for the one loop characteristic functions $\overline{x}(t)$, $\overline{y}(t)$, $\overline{z}(t)$ appear
in \cite{14}.)

With these CW RG functions we can compute $V_{NLL}$ using eq.~\eqref{neweq64} in the model we are considering.

\section{Application to the Standard Model}

We now show how the results of the previous two sections can be applied to the standard model in order to estimate the mass of the Higgs
Boson.  We only consider the case in which there is a single Higgs doublet with no classical mass term.

As was pointed out in \cite{1,2}, there are three things to consider.  First of all, we have the CW renormalization conditions of eqs.~(\ref{eq2}, \ref{eq18}).  Next
there is the stability condition
\begin{equation}
\frac{d}{d\phi} V(\phi = \mu) = 0.
\label{eq112}
\end{equation}
This means that we identify $\mu$ with the vacuum expectation value of $\phi$, that is $\mu=2^{-1/4}G_F^{-1/2}$.  Once these two requirements are satisfied,
we can compute the Higgs mass by the formula
\begin{equation}
m_H^2 = \frac{d^2V(\phi = \mu)}{d\phi^2}/Z(\phi =\mu).
\label{eq113}
\end{equation}
With the renormalization condition of eq.~(18) this just reduces to
\begin{equation}
m_H^2 = \frac{d^2V(\phi = \mu)}{d\phi^2}.
\label{eq114}
\end{equation}

If $V$ is expanded in the form
\begin{equation}
V = \sum_{p=0}^\infty V_{N^pLL}
\label{eq115}
\end{equation}
where $V_{N^pLL}$ is the $N^pLL$ contribution to $V$, then we begin by estimating $V$ by
\begin{equation}
V_m = \sum_{p=0}^m V_{N^pLL} + \pi^2 K_m\phi^4.
\label{eq116}
\end{equation}

The term $\pi^2K_m\phi^4$ in eq.~\eqref{eq116} represents the parts of $V$ coming from those terms in eq.~\eqref{eq115} beyond $N^mLL$
which can be determined by imposing eq.~\eqref{eq2} --- the renormalization condition.
As is discussed in section two above, $V_{N^pLL}$ can be determined in terms of the CW RG functions if they are known to $p + 1$ loop order.  From
section two then, $V_{LL}$ can be found using all five couplings $(x, y, z, r, s)$, $V_{NLL}$ can be found using the three couplings $(x, y, z)$ and
finally $V_{N^2LL}$, $V_{N^3LL}$ and $V_{N^4LL}$ can be found using the single coupling $y$.

The role of $K_m$ in eq.~\eqref{eq116} is to ensure that the CW renormalization condition of eq.~\eqref{eq2} is satisfied.  It is a ``counter-term"; more explicitly in terms of the quantities
$p_n^k$ introduced in eq.~\eqref{eq25} (or the generalization of this expression to accommodate more than two couplings)
\begin{equation}
K_m = \sum_{n=m+2}^\infty p_n^0.
\label{eq117}
\end{equation}
Eqs.~\eqref{eq12} and \eqref{eq27} on their own only ensure that eq.~\eqref{eq2} is satisfied up to a finite order $m$ in the coupling constant expansion; the inclusion of the counter-term ensures that eq.~\eqref{eq2} is satisfied to all orders.
Once expressions for $V_{LL} \ldots V_{N^mLL}$ have been given in terms of the appropriate CW RG functions, there are still two unknowns: the counter-term
$K_m$ and the quartic scalar coupling $y$.  These two are fixed by conditions \eqref{eq2} and \eqref{eq112}, then $V_m$ is used in conjunction with eq.~\eqref{eq114} to estimate $m_H^2$.

More explicitly, $V_{LL}$ is given by eq.~\eqref{eq38} with eq.~\eqref{eq29} leading to
\begin{equation}
V_{LL} = \pi^2 p_1^0 \left(\overline{x}\left(\frac{L}{2}\right), \overline{y}\left(\frac{L}{2}\right),\overline{z}\left(\frac{L}{2}\right),\overline{r}\left(\frac{L}{2}\right),
\overline{s}\left(\frac{L}{2}\right)\right)
\exp\left[-4 \int_0^{L/2} d\tau \gamma_1(\overline{x}(\tau), \ldots , \overline{s}(\tau))\right]\phi^4.
\label{eq118}
\end{equation}
We see by eq.~\eqref{eq27}, $p_1^0 = y$ and by eqs.~(\ref{eq79}, \ref{eq95}), $\gamma_1 = \frac{3}{4} x - \frac{9}{16} r - \frac{3}{16} s$.

When one computes derivatives of the characteristic functions $\overline{x}(t) \ldots \overline{s}(t)$ when evaluating $V_{LL}^{\prime}$, $V_{LL}^{\prime\prime}$ and $V_{LL}^{\prime\prime\prime\prime}$ as required
by eqs.~(\ref{eq2}, \ref{eq112}, \ref{eq113}), the one loop contributions to $\tilde{\beta}^x \ldots \tilde{\beta}^s$ in
eqs.~(\ref{eq74} -- \ref{eq78}) are to be used as at one loop order the CW and MS RG functions
are the same.

For $V_{NLL}$ we need RG functions in the CW renormalization scheme to two loop order.  These are given by eqs.~(\ref{eq108} -- \ref{eq110}) for the limiting case in which the standard model
with only the three couplings $(x, y, z)$ is being considered.  These are used in conjunction with $V_{NLL}$ in eq.~\eqref{neweq64}.  In this equation, we have
\begin{equation}
w_1^0\left( \overline{x}\left(\frac{L}{2}\right), \overline{y}\left(\frac{L}{2}\right),\overline{z}\left(\frac{L}{2}\right)  \right)
=  \overline{y}\left(\frac{L}{2}\right)\exp\left[-4 \int_0^{L/2} d\tau \left(\frac{3}{4} \overline{x}(\tau)\right)\right]
\label{eq119}
\end{equation}
and since by eqs.~(\ref{neweq40}, \ref{eq108}--\ref{eq111})
\begin{equation}
p_1^0 = y \quad \quad p_2^1 = 3y^2 - \frac{3}{4} x^2 \quad \quad p_2^0 = - \frac{25}{6} p_2^1
\label{eq120}
\end{equation}
we also have
\begin{equation}
w_2^0 = \left[ - \frac{25}{2} \overline{y}^2 \left(\frac{L}{2}\right) + \frac{25}{8} \overline{x}^2\left(\frac{L}{2}\right)\right] \exp
\left[-4 \int_0^{L/2} d\tau \left(\frac{3}{4} \overline{x}(\tau)\right)\right].
\label{eq121}
\end{equation}
For consistency, the derivatives of $\overline{x}(t), \overline{y}(t), \overline{z}(t)$ that arise when computing
$V_{NLL}^{\prime}$, $V_{NLL}^{\prime\prime}$ and $V_{NLL}^{\prime\prime\prime\prime}$ are given by the one loop contributions to $\beta^x$, $\beta^y$, $\beta^z$ occurring in eqs.~(\ref{eq108} -- \ref{eq110}).

Finally, for $V_{N^2LL}$, $V_{N^3LL}$ and $V_{N^4LL}$ we have at our disposal only the CW RG functions associated with the single scalar coupling $y$.  These RG
functions are found by combining eqs.~(\ref{eq68}, \ref{eq69}, \ref{eq84}, \ref{eq85}).  Using them, the functions $S_2 \ldots S_4$ appearing in eq.~\eqref{eq8} are given by
\begin{gather}
S_2(\xi) = \frac{1}{4w} + \left(-\frac{175}{16} + \frac{1}{16} \ln |w| - \frac{21}{2} \zeta (3)\right) \frac{1}{w^2} + \left( \frac{1}{16} \ln^2|w| - \frac{103}{16} \ln |w| +\frac{3591}{16} + \frac{21}{2} \zeta (3)\right) \frac{1}{w^3}
\label{eq122}
\\
\begin{split}
S_3(\xi) =& \left(-\frac{7}{8}\zeta(3) - \frac{1}{96}\right) \frac{1}{w} + \left( -\frac{7\pi^4}{40} + \frac{1}{16} \ln |w| + \frac{365}{4} \zeta (5)
+ \frac{1205}{64}  + \frac{239}{8} \zeta(3) \right)  \frac{1}{w^2} \\
&\quad + \left(\frac{16363}{64} + \frac{\ln^2|w|}{64} - \frac{21}{4} \zeta(3) \ln |w| + 273 \zeta(3)   - \frac{351}{64} \ln |w|\right) \frac{1}{w^3}  \\
&\quad+ \left(\frac{7\pi^4}{40} + \frac{1}{64} \ln^3 |w| - \frac{239263}{48} - \frac{1733}{4}\zeta(3)  + \frac{2719}{16} \ln |w| - \frac{311}{128} \ln^2 |w| + \frac{63}{8} \zeta(3) \ln |w| - \frac{365}{4} \zeta (5)\right) \frac{1}{w^4}
\end{split}
\label{eq123}
\end{gather}
and
\begin{equation}
\begin{split}
\hspace{-8cm}S_4(\xi) &= \left(-\frac{7\pi^4}{160} + \frac{45}{8}\zeta(3) - \frac{713}{768} + \frac{365}{32} \zeta(5)\right) \frac{1}{w} + \Biggl[ \frac{365\pi^6}{1008} -
\frac{\ln |w|}{384} - \frac{3449}{6}  \zeta (5) \Biggr. \\
&\quad\qquad\qquad \Biggl.  - \frac{4421}{24} \zeta(3) + \frac{139}{8} \zeta^2(3) - \frac{36897}{32}  \zeta(7) - \frac{7}{32} \zeta(3) \ln |w|
-\frac{5347}{48} + \frac{337\pi^4}{320}\Biggr] \frac{1}{w^2}  \\
&\quad + \Biggl[ -\frac{19325}{32} \zeta(3)  - \frac{37595}{16} \zeta (5)
- \frac{115387}{256} + \frac{365}{8} \ln |w| \zeta (5) + \frac{441}{4} \zeta^2(3) + \frac{1203}{128} \ln |w|\Biggr. \\
&\quad\qquad \Biggl. + \frac{1}{64} \ln^2|w| + \frac{239}{16} \ln |w| \zeta(3) + \frac{721\pi^4}{160} - \frac{7\pi^4}{80} \ln |w|\Biggr] \frac{1}{w^3} \\
&\quad +\Biggl[ - \frac{63}{32} \ln^2|w| \zeta (3) - \frac{1250731}{192} + \frac{1545}{8} \ln |w| + \frac{1}{256} \ln^3|w|
-\frac{1323}{4} \zeta^2(3)\Biggr. \\
&\quad\qquad \Biggl.+ \frac{3297}{16} \ln |w|\zeta(3) - \frac{1055}{512} \ln^2 |w| - \frac{365}{16} \zeta (5)
- \frac{119837}{16} \zeta (3) + \frac{7\pi^4}{160}\Biggr] \frac{1}{w^4} \\
& \quad+ \Biggl[ - \frac{365\pi^6}{1008} - \frac{3179\pi^4}{320} + \frac{7\pi^4}{40} \ln |w|
+ \frac{51712991}{384} + \frac{1625}{8} \zeta^2(3) + \frac{1}{256} \ln^4|w| - \frac{13927}{32} \ln |w|\zeta(3)\Biggr. \\
&\qquad\quad + \frac{36897}{32} \zeta(7) + \frac{1505921}{96} \zeta(3) - \frac{965209}{192} \ln |w| + \frac{500849}{96} \zeta (5) \\
&\quad \qquad\Biggl. - \frac{625}{768} \ln^3|w| + \frac{63}{16} \ln^2|w| \zeta(3) + \frac{43815}{512} \ln^2|w| - \frac{365}{4} \ln |w| \zeta (5)\Biggr] \frac{1}{w^5}.
\end{split}
\label{eq124}
\end{equation}
With one coupling, we have $V_{N^pLL} = \pi^2 y^{p+1} S_p(yL)\phi^4$ for $p = 2, 3, 4$.

It is now possible to implement our program for determining the mass of the Higgs. This requires knowledge of  $x, z, r$ and $s$ at the mass scale $v$.
The couplings $x, z, r$ and $s$ are defined in terms of the Yukawa and gauge couplings $g_t,\ g_3,\ g_2$ and $g_1$ by eqs.~(\ref{eq70} -- \ref{eq73}). These in turn are related to the measured quantities $m_t$ (the top quark mass), $\theta_w$ (the weak angle), $M_W$ (the $W$- Boson mass), $\alpha_s$ (the strong structure constant) and $\alpha$ (the fine structure constant), all of which are known at the mass scale set by the $Z$-Boson. These relations are
\begin{align}
x_0 &= \frac{\alpha}{2\pi} \left(\frac{m_t}{M_W\sin\theta_w}\right)^2
\label{eq125}
\\
z_0 &= \frac{\alpha_s}{\pi}
\label{eq126}
\\
r_0 &= \frac{\alpha}{\pi\sin^2\theta_w}
\label{eq127}
\\
s_0 &= \frac{\alpha}{\pi\cos^2\theta_w}.
\label{eq128}
\end{align}
where the subscript $0$ means that these are evaluated at the mass of the $Z$-Boson.
From the Particle Data Group \cite{15}, at the mass of the $Z$-Boson (91.1876 $GeV/c^2$), $\alpha = 1/128.91$, $\alpha_s = .1176$, $\sin^2\theta_w = .23119$, $M_w = 80.398\, GeV/c^2$ and
$m_t = 171.3\, GeV/c^2$. It is now necessary
to evaluate these couplings at the vacuum expectation value $v = 2^{-1/4}G_F^{-1/2}$ (taking $G_F$ to be $1.16637 \times 10^{-5}\,\left (GeV^/c^2 \right )^{-2}$). To do this,
we use the one loop limit of the RG equations that follow from eqs.~(\ref{eq74} -- \ref{eq78}) as a suitable approximation
\begin{align}
\mu \frac{dx}{d\mu} &= \frac{9}{4} x^2 - 4xz
\label{eq129}
\\
\mu \frac{dz}{d\mu} &= -\frac{7}{2} z^2
\label{eq130}
\\
\mu \frac{dr}{d\mu} &= -\frac{19}{12} r^2
\label{eq131}
\\
\mu \frac{ds}{d\mu} &= \frac{41}{12} s^2.
\label{eq132}
\end{align}
Eqs.~(\ref{eq130} -- \ref{eq132}) have solutions \cite{14}
\begin{align}
z &= \frac{z_0}{1 + \frac{7}{2} z_0\ln \left(\frac{\mu}{\mu_0}\right)}
\label{eq133}
\\
r &= \frac{r_0}{1 + \frac{19}{12} r_0\ln \left(\frac{\mu}{\mu_0}\right)}
\label{eq134}
\\
s &= \frac{s_0}{1 - \frac{41}{12} s_0\ln \left(\frac{\mu}{\mu_0}\right)}.
\label{eq135}
\end{align}
Dividing eq.~\eqref{eq129} by eq.~\eqref{eq130} leads to the homogeneous equation
\begin{equation}
\frac{dx}{dz} = - \frac{9}{14} \left(\frac{x}{z}\right)^2 + \frac{8}{4} \left(\frac{x}{z}\right)
\label{eq136}
\end{equation}
whose solution is
\begin{equation}
x = \frac{(2/9)z}{1-[(1-2/9(z_0/x_0)](z/z_0)^{-1/7}}
\label{eq137}
\end{equation}
Using $(x_0, z_0, r_0, s_0,)$ given by eqs.~(\ref{eq125} -- \ref{eq128}) at the mass scale $\mu_0=91.1876 \ GeV/c^2$ then
eqs.~(\ref{eq133} -- \ref{eq135}, \ref{eq137}) yield $(x, z, r, s)$ at the mass scale $\mu=v=2^{-1/4}G_F^{-1/2}$.

We can now proceed to compute the Higgs mass at each order of the expansion of $V$ in the $N^pLL$ expanion.  With eq.~\eqref{eq116} for $V_m$, we use eq.~\eqref{eq2} to fix $K_m$ in terms of $y$ and then use
eq.~\eqref{eq112} to solve for $y$ itself.  In this paper the only acceptable values for $y$ are positive in order to ensure physical stability of the theory for reasonable values of $\phi^2$, as will be discussed below.  With these values of $y$ (and $K_m$)
eq.~\eqref{eq116} can be used give an explicit expression for $V_m$.  Eq.~\eqref{eq114} can then be used to evaluate $m_H^2$.  Only real and positive values of $m_H^2$ are acceptable.
We note that it is not necessary to find explicit results for the integrals and running couplings appearing in eqs.~(\ref{eq118}, \ref{eq119}, \ref{eq121}). The derivatives of these expressions at $\phi=v$
that are needed to evaluate the Higgs mass are determined completely in terms of the RG functions and boundary values at $\phi=v$. Thus our methodology can be applied to very complicated models and is an important tool in its own right.

We present, in Table 1, the values of $K_m$, $\lambda = \pi^2 y$, $m_H$ for $m = 0, 1, 2, 3, 4$ when $(x, y, z, r, s)$ contribute at $LL$ order, $(x, y, z)$ contribute at $NLL$ order and only $y$ contributes beyond that. (The units for $m_H$ are $GeV/c^2$.) It is important to emphasize that the values for $K_m$ listed in Table 1 arise because of the functional dependence of $K_m$ on the coupling $y$; first $K_m$ is expressed in terms of $y$ by using eq.~\eqref{eq2} and then $y$ is fixed by eq.~\eqref{eq112}.

\begin{table}[ht]
\centering
\begin{tabular}{|c|c|c|c|}
\hline
$m$ &$K_m$ & $\lambda$ & $m_H$\\
		\hline
		0 & -.0586 & .536 & 219\\
		1 &       &      &       \\
		2 & -.0431 & .439 & 188 \\
		3 &       &      &       \\
		4 & -.0346 & .363 & 163 \\
		\hline
\end{tabular}
\caption{Calculated results for the standard model to three significant digits.}
\end{table}

No entry occurs for $m = 1$ or $m = 3$ as the values of $y$ that follow from $V_1$ and $V_3$ are  negative  and
unacceptable.  This appears to be due to the large negative contribution to $S_1$ and $S_3$ coming from terms of order $\frac{1}{w^2}$ and
$\frac{1}{w^4}$ respectively.

\begin{figure}[htb]
\centering
\includegraphics[scale=0.5]{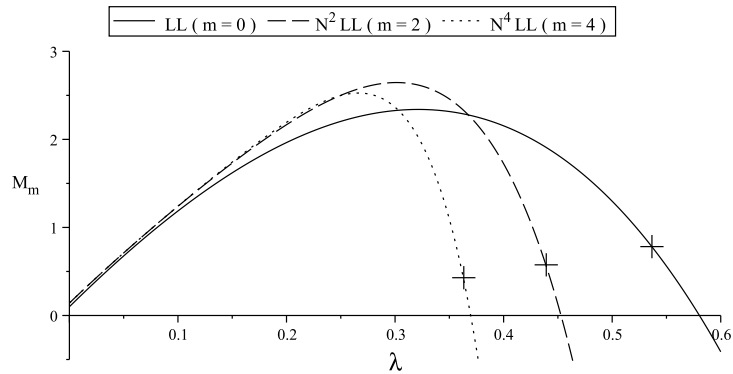}
\caption{
The dimensionless ratio $M_m=\left.\frac{1}{v^2}\frac{d^2}{d\phi^2}V_m\right\vert_{\phi=v}$ plotted  as a function of $\lambda$.
}
\label{mn_fig}
\end{figure}

\begin{figure}[htb]
\centering
\includegraphics[scale=0.8]{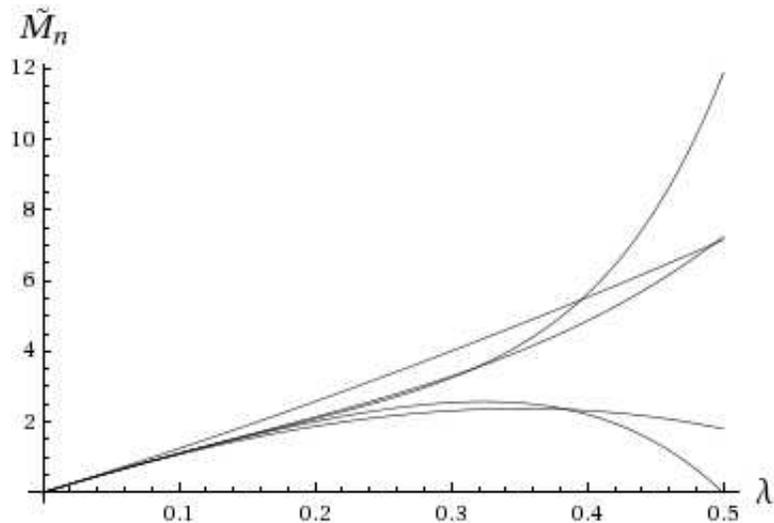}
\caption{The dimensionless quantity $\tilde M_n=\left.\frac{1}{v^2}\frac{d^2\left(V_n-\pi^2K_n\phi^4\right)}{d\phi^2}\right\vert_{\phi=v}$ is plotted as a function of $\lambda$ for the $O(4)$ scalar theory. The upper curves represent the  even orders ($n=0,2,4$) and the lower curves represent the odd orders ($n=1,3$).}
\label{mn_fig2}
\end{figure}

The second derivative of the order $m$ estimate for the effective potential, normalized to the scale $v^2$,
\begin{equation}
M_m=\left.\frac{1}{v^2}\frac{d^2}{d\phi^2}V_m\right\vert_{\phi=v}
\end{equation}
can be viewed as a function of the scalar field coupling $\lambda$ once the counter-term $K_m$ has been expressed in terms of $\lambda$. In Figure \ref{mn_fig} we present curves for the dimensionless quantity $M_m (m=0,2,4)$ for positive values of $\lambda$ while $M_m$ is positive.
The crosses on the curves correspond to the values of $\lambda$ and $m_H$ found by our approach and listed in Table 1 for $m=(0,2,4)$. Table 1 and Figure \ref{mn_fig} suggest a tendency for both $\lambda$ and $m_H$ to decrease with increasing order $m$.
We can gain further insight on this trend in the $O(4)$ scalar theory by extracting the  counter-term  from the second derivative, normalized to the scale $v^2$,
\begin{equation}
\tilde M_n=\left.\frac{1}{v^2}\frac{d^2\left(V_n-\pi^2K_n\phi^4\right)}{d\phi^2}\right\vert_{\phi=v}~. 
\end{equation}
For the pure scalar field theory case the resulting dimensionless expressions are shown as a function of $\lambda$ in Figure \ref{mn_fig2}. One can see the distinction between even and odd orders in the Figure, and one can also see evidence of slow convergence towards a result which would lie between the even and odd envelopes of the curves. Because $\tilde M_n$ represents the field-theoretical (i.e., counter-term-independent) contributions to the Higgs mass, it is evident that even orders provide an upper bound  on $m_H$ and odd orders provide a lower bound on $m_H$. Although the lower bound is trivial (i.e., $m_H=0$), this does not obviate the interpretation of $m_H$  at odd orders as an upper bound.

\section{Discussion}

In this paper we have presented a systematic way of using the RG equation to sum all of the logarithms contributing to $V$ at order $N^pLL$ in terms of the $(p + 1)$ order RG functions, provided we use the CW renormalization scheme and have only one form of logarithm (here $L = \log\left[\phi^2/\mu^2\right]$) contributing to $V$. We have applied our method of analysis to the conformal limit of the standard model with a single scalar field, as was originally envisaged by Coleman and Weinberg \cite{4}. This has led to a surprisingly interesting sequence of estimates for the Higgs mass and the quartic scalar couplings.

It was not anticipated that the improvements to the approach, originally used in \cite{1,2}, introduced in this paper and \cite{3} would lead to a sequence of decreasing estimates for the Higgs mass as listed in Table 1 above.  The values of these estimates suggest that increasing the order $m$ to 6 and beyond (if that were feasible) would lead to Higgs mass estimates closer to the generally expected range of possible values.  A compilation of predictions of the Higgs mass in different scenarios is given in
ref.~\cite{19}, and a discussion on its limits is given in ref.~\cite{16}.  In our approach we have made use of all known RG functions relevant to any part of the standard model.  To make further progress using this approach will require knowledge of RG functions at a higher loop order than is currently available.

Even though we have not come up with a definitive prediction of the Higgs mass within the standard model, we feel that our results establish the viability of the Coleman-Weinberg mechanism to  generate spontaneous symmetry breaking and to provide a mass for the Higgs scalar particle.  We have done this by the use of the RG-improved effective potential. We propose that the masses generated in Section 4 above be viewed as a decreasing sequence of upper bounds on the actual Higgs mass in the standard model.

A significant insight into the standard model effective potential can be gained by applying our method of analysis to a simplified pure $O(4)$ scalar field theory obtained from the standard model by setting all couplings except $\lambda=\pi^2y$ to zero.  We present in Table 2 the results for $K_m$, $\lambda$ and $m_H$ in this simplified model using exactly the same steps as were used to derive the results in Table 1 for the standard model.

\begin{table}[ht]
\centering
\begin{tabular}{|c|c|c|c|}
\hline
$m$ &$K_m$ & $\lambda$ & $m_H$\\
		\hline
		0 & -.0585 & .534 & 221\\
		1 &   0    &  0    &  0     \\
		2 & -.0390 & .417 & 186 \\
		3 &   0    &   0   &   0    \\
		4 & -.0321 & .354 & 165 \\
		\hline
\end{tabular}
\caption{Calculated results for the $O(4)$ scalar theory to three significant digits.}

\end{table}

\begin{figure}[htb]
\centering
\includegraphics[scale=0.5]{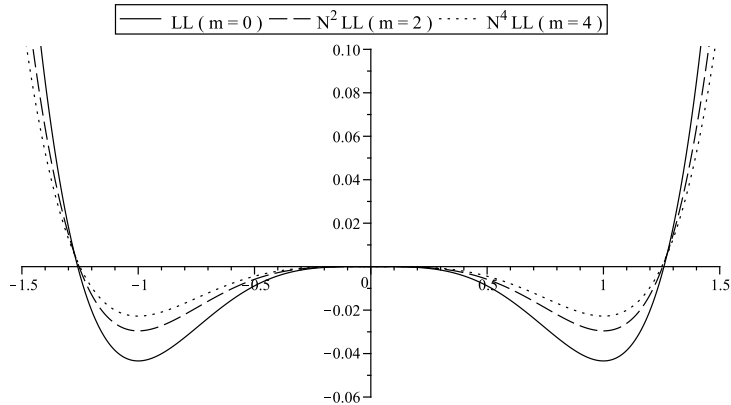}
\caption{$V_m$ is plotted as a function of $\phi/v$ with $\lambda$ as in Table 2.}.
\label{Veff_fig}
\end{figure}

The similarity between the results of Tables 1 and 2 indicates that $y$ is the dominant coupling in these considerations, much more than $x$, $z$, $r$ or $s$.  We note the vanishing values for $K_m$, $\lambda$ and $m_H$ in Table 2 for $m = 1, 3$.  For this simplified model our method yields the acceptable but trivial solution $\lambda = 0$ for all values of $m$.  In Table 2 we only include the non-trivial solutions for $m = 0, 2, 4$.
For these non-trivial solutions we can plot $V_m$ as a function of $\phi$ for values of $\phi$ near the VeV scale $v$, something which cannot be easily done in the standard model.  This plot is provided in Figure \ref{Veff_fig}.

Remarkably, the plots of $V_0$, $V_2$ and $V_4$ have the well known shape of a spontaneous symmetry breaking potential when restricted to $\phi$ values near the location of the minimum.  These potentials also have a singularity at $\phi = \pm v \exp \left( \pi^2/6\lambda \right) $ ( \textit{i.e.} when $w = 0$ ). This is significantly far from the region near the minimum.

\begin{figure}[htb]
\begin{minipage}[b]{0.45\linewidth}\centering
\includegraphics[scale=0.35]{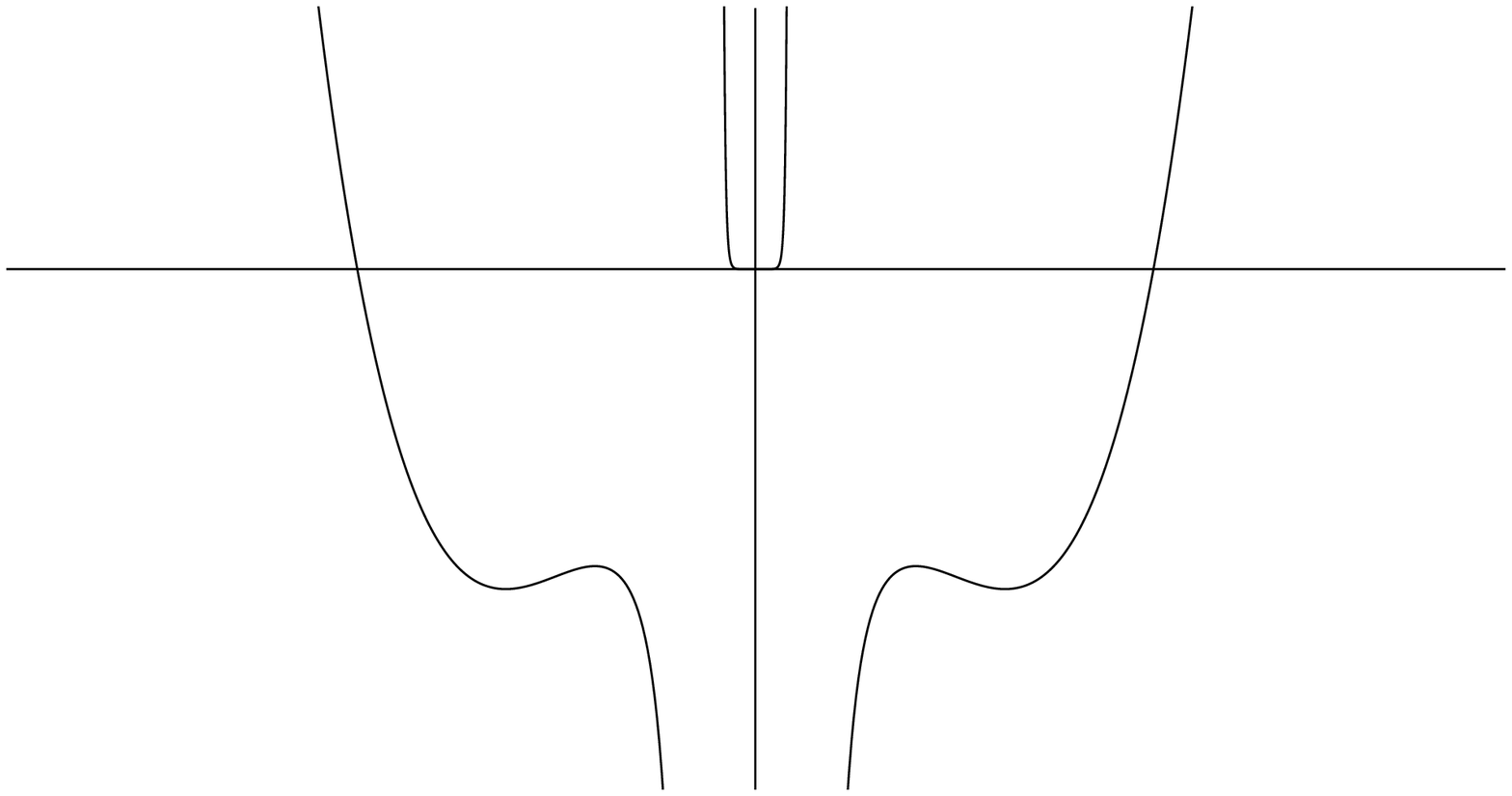}
\caption{Shape of $V_m$ for $\lambda<0$ as a function of $\phi$ for $m = 0, 2, 4$.}.
\label{Veff_fig_2}
\end{minipage}
\hspace{0.5cm}
\begin{minipage}[b]{0.45\linewidth}\centering
\includegraphics[scale=0.35]{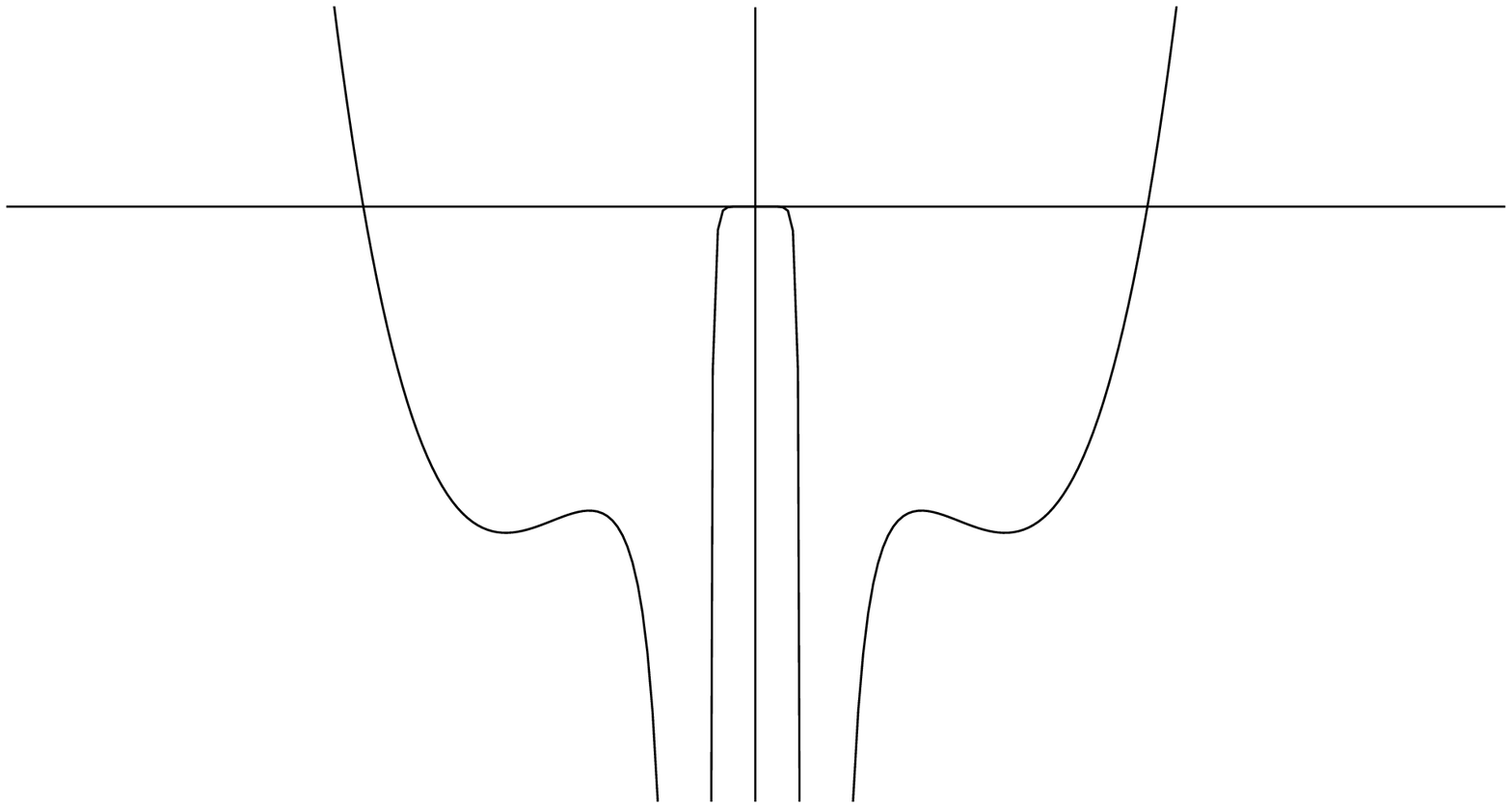}
\caption{Shape of $V_m$ for $\lambda<0$ as a function of $\phi$ for $m = 1, 3$.}.
\label{Veff_fig_3}
\end{minipage}
\end{figure}

In addition to the positive and zero $\lambda$-solutions in the pure $O(4)$ scalar field model referred to above, there are negative $\lambda$-solutions.  We have heretofore rejected negative $\lambda$-solutions as unacceptable.  In contrast to the standard model, in the $O(4)$ model we can plot $V_m$ as a function of $\phi$ with these negative values of $\lambda$.  We show the shape of $V_m(\phi)$ for the appropriate negative $\lambda$-values for $m = 0, 2, 4$ in Figure \ref{Veff_fig_2} and for $m = 1, 3$ in Figure \ref{Veff_fig_3}.

For the even $m$ cases $(m = 0, 2, 4)$ we note the existence of a tightly bound minimum at $\phi = 0$, singularities at $|\phi| < v$ (since $\lambda < 0$) and local minima at $\phi = \pm v$.  On the other hand, for the odd $m$ cases $(m = 1, 3)$ we note the existence of a highly unstable maximum at $\phi = 0$, singularities at $|\phi| < v$ (since $\lambda < 0$) and local minima at $\phi = \pm v$.
The occurrence of a singularity at $w = 0$ in $V_m$ may be considered pathological but away from the singular points the form of $V_m$  is interesting.  Whether this feature has a role to play in the standard model is an open question which may be worth pursuing. It has been shown \cite{3} in the scalar model that summing portions of the contributions to $V_m$ beyond order $m = 4$ may shift such singularities.

We have attempted setting $K_m = 0$ in eq.~\eqref{eq116}, and then determining the single remaining unknown $y$ by using either eq.~\eqref{eq2} or eq.~\eqref{eq112}.  Neither
of these attempts leads to acceptable values of $y$ or $m_H^2$; one must employ the counter-term $K_m$ in eq.~\eqref{eq116} to get reasonable values for these parameters at any value of $m$. In fact, by having introduced the counter-term, we are availing ourselves of information about terms, independent of $L = \log \frac{\phi^2}{\mu^2}$, beyond the $N^pLL$ contribution to $V$.  We have been unable to establish any other viable alternative to the counter-term approach.

Whereas in this paper we have used the CW renormalization scheme, preliminary investigations indicate that it may be possible to adapt our approach to incorporate the MS renormalization scheme, at least in the single coupling $O(4)$ scalar model.
Using the MS renormalization scheme to compute the $LL$ and $NLL$ contributions to $V$ when there is only the coupling $y$, realistic values of $m_{H}^{2}$ and $y$ follow from eqs.~\eqref{eq112} and \eqref{eq114} only if the counter-term $K_{m}$ of eq.~\eqref{eq116} is included and the condition of eq.~\eqref{eq2} is applied.
Strictly speaking, eq.~\eqref{eq2} is not part of the MS renormalization scheme, though it might possibly be used to fix the physical value of $y$ in the MS scheme in a way analogous to using the gap equation to fix a physical mass.

We hope to develop this formalism in several other ways.  First, inclusion of a mass term $-m^2\phi^2$ into the classical action should be considered \cite{23}.
Next, the inclusion of more scalars beyond an $SU(2)$ doublet should be dealt with, as additional scalars are necessary \cite{20} in any
supersymmetric extension of the standard model.  A further problem to be addressed concerns working with summing logarithmic contributions
to $V$ in the standard model using MS RG functions rather than converting them to the CW scheme, even though this would entail having a separate logarithm for each coupling
(see eq.~\eqref{eq86}) and not being able to fix the terms $p_{p+1}^0$ in eq.~\eqref{eq24} by using some analogue of eq.~\eqref{eq27}.  We would also like to see if the RG
methods that have been developed could be employed in the consideration of other physical processes \cite{24}, or the contributions to the
effective action arising due to an external magnetic field \cite{25}.

\section*{Acknowledgements}

This work was largely inspired by the late Victor Elias.  Roger Macleod had a useful suggestion.  There was helpful correspondence with C.~Ford and S.~Martin. NSERC (Natural Science \& Engineering Research Council of Canada)  provided funding for RBM and TGS.

\clearpage
\section*{Appendix 1: Method of Characteristics Solution at $NLL$ and $N^2LL$ Order}
The computation of $V_{NLL}$ begins by noting that by eq.~\eqref{eq28}
\begin{equation}
p_{n+2}^n + \gamma_1 p_{n+1}^n = \frac{1}{2n} \left[\left(\beta_2^x \frac{\partial}{\partial x} + \beta_2^y \frac{\partial}{\partial y} - 4\gamma_1\right)p_{n+1}^{n-1} + \left(\beta_3^x \frac{\partial}{\partial x} + \beta_3^y \frac{\partial}{\partial y} -4\gamma_2\right)p_n^{n-1}\right].
\label{eq39}
\end{equation}
so that together eqs.~(\ref{eq29}, \ref{eq32}, \ref{eq34}, \ref{eq39}) imply that
\begin{equation}
w_{n+2}^n = \frac{1}{2n}\left[\frac{d}{dt} w_{n+1}^{n-1} + D(t) w_n^{n-1}\right]
\label{eq40}
\end{equation}
where
\begin{equation}
D(t) = -\gamma_1 \left(\beta_2^x \frac{\partial}{\partial \overline{x}} + \beta_2^y \frac{\partial}{\partial \overline{y}} - 4\gamma_1\right) + \left(\beta_3^x \frac{\partial}{\partial \overline{x}} + \beta_3^y \frac{\partial}{\partial \overline{y}} - 4\gamma_2\right).
\label{eq41}
\end{equation}
Iterating eq.~\eqref{eq40} shows that
\begin{equation}
w_{n+2}^n = \frac{1}{2^nn!} \left[ \frac{d^n}{dt^n} w_2^0 + \left( \frac{d^{n-1}}{dt^{n-1}} D(t) + \frac{d^{n-2}}{dt^{n-2}} D(t) \frac{d}{dt} + . . . + D(t) \frac{d^{n-1}}{dt^{n-1}}\right) w_1^0 \right] .
\label{eq42}
\end{equation}
One can inductively prove the identity
\begin{equation}
\left( \frac{d^{n-1}}{dt^{n-1}} f + \frac{d^{n-2}}{dt^{n-2}} f \frac{d}{dt} + \ldots + \frac{d}{dt} f \frac{d^{n-2}}{dt^{n-2}} + f \frac{d^{n-1}}{dt^{n-1}}\right)g  = \frac{d^n}{dt^n} (\phi g) - \phi \frac{d^n}{dt^n} g \qquad (\ \frac{d\phi}{dt} \equiv f \ ).
\label{eq43}
\end{equation}

To employ eq.~\eqref{eq43} to simplify eq.~\eqref{eq42} we need to commute the functional derivatives appearing in $D(t)$ (see eq.~\eqref{eq41}) through $\frac{d}{dt}$ so that they act
on $g$ before $\frac{d}{dt}$ does.  (This step was not considered properly in eq.~(B22) of ref.~\cite{3}.)  In order to do this, we first write $D(t)$ in eq.~\eqref{eq41} in the form
\begin{equation}
D(t) = A^i \frac{\partial}{\partial \overline{x}^i(t)} + B
\label{eq44}
\end{equation}
where
\begin{gather}
\overline{x}^1(t) \equiv \overline{x}(t),
\label{eq45}
\\
\overline{x}^2(t) \equiv \overline{y}(t),
\label{eq46}
\\
A^1(\overline{x}^i(t)) \equiv -\gamma_1 \beta_2^x + \beta_3^x,
\label{eq47}
\\
A^2(\overline{x}^i(t))\equiv -\gamma_1 \beta_2^y + \beta_3^y,
\label{eq48}
\end{gather}
and
\begin{equation}
B(\overline{x}^i(t)) \equiv 4(\gamma_1^2 - \gamma_2).
\label{eq49}
\end{equation}
Furthermore, using eqs.~(\ref{eq30}, \ref{eq31}),
\begin{equation}
\frac{d}{dt}
= \beta_2^x (\overline{x}(t), \overline{y}(t))\frac{\partial}{\partial\overline{x}(t)} + \beta_2^y(\overline{x}(t), \overline{y}(t))
\frac{\partial}{\partial\overline{y}(t)} + \frac{\partial}{\partial t}
\equiv \Lambda^i \frac{\partial}{\partial\overline{x}^i} + \frac{\partial}{\partial t}
.
\label{eq50}
\end{equation}
We now note that
\begin{equation}
\begin{split}
A^i \frac{\partial}{\partial\overline{x}^i} \frac{df}{dt} &= A^i \frac{\partial}{\partial\overline{x}^i}
\left(\Lambda^j \frac{\partial}{\partial\overline{x}^j} + \frac{\partial}{\partial t}\right) f
= A^i \left[ \left( \Lambda^j \frac{\partial}{\partial\overline{x}^j} + \frac{\partial}{\partial t}\right)
\frac{\partial f}{\partial\overline{x}^i}
+ \frac{\partial \Lambda^j}{\partial\overline{x}^i}\frac{\partial f}{\partial\overline{x}^j}\right]
\\
&=  A^i \left[ \frac{d}{dt} \delta_{ij} + (\mathbf{M})_{ij} \right]\frac{\partial f}{\partial\overline{x}^j}
\end{split}
\label{eqblank50}
\end{equation}
where
\begin{equation}
(\mathbf{M})_{ij} = \frac{\partial\Lambda_j}{\partial \overline{x}^i},
\label{eq51}
\end{equation}
and so by iterating we obtain
\begin{equation}
A^i \frac{\partial}{\partial\overline{x}^i} \left(\frac{d}{dt}\right)^pf =
A^i \left[ \left( \frac{d}{dt} + \mathbf{M}\right)^p\right]_{ij}
\frac{\partial f}{\partial\overline{x}^j} .
\label{eq52}
\end{equation}
If we now define
\begin{equation}
(\mathbf{U}(t,0))_{ij} = \delta_{ij} + \sum_{n=1}^\infty \int_0^t d\tau_1 \int_0^{\tau_1} d\tau_2 \ldots
\int_0^{\tau_{n-1}}d\tau_{n}
\left[ \ \mathbf{M}(\tau_n)\mathbf{M}(\tau_{n-1})...\mathbf{M}(\tau_2) \mathbf{M}(\tau_1) \ \right]_{ij}.
\label{eq53}
\end{equation}
then it is evident that
\begin{equation}
\frac{d}{dt}(\mathbf{U} (t,0)f) = \mathbf{U} (t,0) \left(\frac{d}{dt} + \mathbf{M}\right) f
\label{eq54}
\end{equation}
and that
\begin{equation}
\mathbf{U}^{-1}(t,0) = \mathbf{U}(0,t) = 1 + \sum_{n=1}^\infty (-1)^n \int_0^t d\tau_1 \ldots \int_0^{\tau_{n-1}} d\tau_n
\left[ \  \mathbf{M}(\tau_1) \ldots  \mathbf{M}(\tau_n) \  \right]
\label{eq55}
\end{equation}
(An operator analogous to $\mathbf{U}$ arises in standard perturbation theory.)  Together, eqs.~(\ref{eq51}--\ref{eq55}) show that
\begin{equation}
A^i \frac{\partial}{\partial\overline{x}^i} \left(\frac{d}{dt}\right)^p f = A^i \left[ \mathbf{U}(0,t)\left(\frac{d}{dt}\right)^p
\mathbf{U}(t,0)\right]_{ij}\frac{\partial}{\partial\overline{x}_j}f.
\label{eq56}
\end{equation}
We now find that by eqs.~(\ref{eq43}, \ref{eq44}, \ref{eq56})
\begin{align}
& \left(\frac{d^{n-1}}{dt^{n-1}} D(t) + \frac{d^{n-2}}{dt^{n-2}} D(t) \frac{d}{dt} + \ldots + \frac{d}{dt} D(t)
\frac{d^{n-2}}{dt^{n-2}} + D(t) \frac{d^{n-1}}{dt^{n-1}}\right)w_1^0(\overline{x}^i(t),t)\nonumber \\
&\qquad = \frac{d^n}{dt^n}\left(\tilde{Z}_j(t)\zeta_{1j}^0(\overline{x}^i(t),t)\right) -
\tilde{Z}_{j}(t)\frac{d^n}{dt^n}\zeta_{1 j}^0(\overline{x}^i(z),t) + \frac{d^n}{dt^n} \left(\tilde{B}(t) w_1^0(\overline{x}^i(t),t)\right) - \tilde{B}(t) \frac{d^n}{dt^n} w_1^0 (\overline{x}^i(t),t),
\label{eq57}
\end{align}
where
\begin{gather}
\tilde{Z}_j(t) \equiv \left(\int_0^t d\tau\,A^i(\overline{x}^i(\tau))\mathbf{U}_{ij}(0,\tau)\right)
\label{eq58}
\\
\tilde{\zeta}_{1j}^0(\overline{x}^i(t),t) \equiv \mathbf{U}_{jk}(t,0) \frac{\partial}{\partial\overline{x}^k(t)}
w_1^0 (\overline{x}^i(t),t)
\label{eq59}
\end{gather}
and
\begin{equation}
\tilde{B}(t) = \int_0^t d\tau \; B(\overline{x}^i(\tau)).
\label{eq60}
\end{equation}
Upon combining eqs.~(\ref{eq35}, \ref{eq42}, \ref{eq57}) we obtain
\begin{equation}
\begin{split}
\overline{V}_{NLL} (\overline{x}^{i} (t),t) &= \pi^2 \phi^4 \sum_{k=0}^\infty \frac{1}{k!}\left(\frac{L}{2}\right)^k\left[ \left(\frac{d}{dt}\right)^k
w_2^0(\overline{x}^i(t),t) + \left(\frac{d}{dt}\right)^k\left(\tilde{Z}_j(t)\zeta_{1j}^0 (\overline{x}^i(t),t)\right)\right.
\\& \left. - \tilde{Z}_j(t)\left(\frac{d}{dt}\right)^k\zeta_{1j}^0(\overline{x}^i(t),t) + \left(\frac{d}{dt}\right)^k\left(\tilde{B}(t)w_{1}^0 (\overline{x}^i(t),t)\right) - \tilde{B}(t)\left(\frac{d}{dt}\right)^kw_{1}^0(\overline{x}^i(t),t)\right].
\end{split}
\label{eq61}
\end{equation}
If we now employ Taylor's theorem with eq.~\eqref{eq61}, it follows that
\begin{equation}
\begin{split}
\overline{V}_{NLL}&= \pi^2\phi^4 \left[w_2^0 \left(\overline{x}^i \left(t + \frac{L}{2}\right), t + \frac{L}{2}\right)  +  \left(\tilde{Z}_j\left(t + \frac{L}{2}\right) - \tilde{Z}_j(t)\right) \zeta_{1j}^0
\left(\overline{x}^i \left(t + \frac{L}{2}\right), t + \frac{L}{2}\right)\right.
\\
& \qquad\quad \left. +  \left(\tilde{B}\left(t + \frac{L}{2}\right) - \tilde{B}(t)\right) w_{1}^0
\left(\overline{x}^i \left(t + \frac{L}{2}\right), t + \frac{L}{2}\right)\right]
\end{split}
\label{eq62}
\end{equation}
and so by eq.~\eqref{eq36}
\begin{equation}
V_{NLL} = \pi^2\phi^4 \left[ w_2^0 \left(\overline{x}^i\left(\frac{L}{2}\right),\frac{L}{2}\right) + \tilde{Z}_j\left(\frac{L}{2}\right)\zeta_{1j}^0 \left(\overline{x}^i
\left(\frac{L}{2}\right), \frac{L}{2}\right) + \tilde{B}\left(\frac{L}{2}\right) w_1^0 \left(\overline{x}^i\left(\frac{L}{2}\right),\frac{L}{2}\right)\right]
\label{eq63}
\end{equation}
or, more explicitly
\begin{equation}
\begin{split}
V_{NLL}&= \pi^2\phi^4 \exp \left[ -4 \int_0^{L/2} d\tau \gamma_1 (\overline{x}^i (\tau)) \right]\Biggl\{ p_2^0 \left(\overline{x}^i\left(\frac{L}{2}\right)\right)  \Biggr.
\\
&\qquad + \int_0^{L/2} d\tau\left[ \left(-\gamma_1(\overline{x}^i(\tau))\beta_2^{x^{i}}(\overline{x}^i(\tau)) + \beta_3^{x^{i}} (\overline{x}^i(\tau))\right)\mathbf{U}_{ij}(0,\tau)\right].\left[ \mathbf{U}_{jk}\left(\frac{L}{2} ,0\right)\frac{\partial}{\partial\overline{x}^k(\frac{L}{2})} p_1^0 \left(\overline{x}^i\left(\frac{L}{2}\right)\right)\right]
\\
&\qquad\qquad\qquad \Biggl. + 4 \int_0^{L/2} d\tau \left[\gamma_1^2 (\overline{x}^i(\tau)) - \gamma_2(\overline{x}^i(\tau))\right] p_1^0\left(\overline{x}^i\left(\frac{L}{2}\right)\right)\Biggr\} .
\end{split}
\label{eq64}
\end{equation}
We have used the fact that $\tilde{B}(0) = 0 = \tilde{Z}_i(0)$.
$V_{N^{2}LL}$ can be computed using the approach used to obtain $V_{NLL}$.  To begin, just as eq.~\eqref{eq39} follows from eq.~\eqref{eq28}, we find that
\begin{equation}
\begin{split}
p_{n+3}^n + \gamma_1 p_{n+2}^n + \gamma_2 p_{n+1}^n &= \frac{1}{2n} \left[ \left( \beta_2^x \frac{\partial}{\partial x} + \beta_2^y \frac{\partial}{\partial y} - 4\gamma_1\right) p_{n+2}^{n-1}
\right.
\\
&\qquad \left.  + \left(\beta_3^x \frac{\partial}{\partial x} + \beta_3^y \frac{\partial}{\partial y} - 4\gamma_2\right) p_{n+1}^{n-1} +
\left(\beta_4^x \frac{\partial}{\partial x} + \beta_4^y \frac{\partial}{\partial y} - 4\gamma_3\right)p_n^{n-1}\right].
\end{split}
\label{eq65}
\end{equation}
With the definitions of eqs.~(\ref{eq29}--\ref{eq31}), we see that eqs.~(\ref{eq34}, \ref{eq40}, \ref{eq65}) together lead to
\begin{equation}
\begin{split}
w_{n+3}^n (\overline{x}(t), \overline{y}(t), t)
&= \frac{1}{2n}\Biggl[ -\gamma_2 \frac{d}{dt} w_n^{n-1} - \gamma_1 \left(\frac{d}{dt} w_{n+1}^{n-1}
+ D(t) w_n^{n-1}\right) \Biggr.
\\
&\quad \Biggl. + \frac{d}{dt} w_{n+2}^{n-1} + \left(\beta_3^x\frac{\partial}{\partial \overline{x}} + \beta_3^y\frac{\partial}{\partial \overline{y}} - 4\gamma_2\right)w_{n+1}^{n-1} + \left(\beta_4^x \frac{\partial}{\partial \overline{x}} + \beta_4^y \frac{\partial}{\partial \overline{y}} - 4\gamma_3\right) w_n^{n-1}\Biggr]
\\
&= \frac{1}{2n} \left[ \frac{d}{dt} w_{n+2}^{n-1} + D(t) w_{n+1}^{n-1} + \Delta(t) w_n^{n-1}\right]
\end{split}
\label{eq66}
\end{equation}
where
\begin{equation}
\Delta (t) = \left[\gamma_1^2 - \gamma_2\right]\left[ \beta_2^x \frac{\partial}{\partial \overline{x}} + \beta_2^y \frac{\partial}{\partial \overline{y}} - 4\gamma_1\right] -\gamma_1 \left[\beta_3^x \frac{\partial}{\partial \overline{x}} + \beta_3^y \frac{\partial}{\partial \overline{y}} - 4\gamma_2\right]
+ \left[\beta_4^x \frac{\partial}{\partial \overline{x}} + \beta_4^y \frac{\partial}{\partial \overline{y}} - 4\gamma_3\right].
\label{eq67}
\end{equation}
Again one can iterate eq.~\eqref{eq66} to obtain $w_{n+3}^n$ in terms of $w_1^0$, $w_2^0$ and $w_3^0$ as well as the two and three loop RG functions in the CW scheme.  The
summations needed to compute $V_{N^{2}LL}$ can then be performed using the same techniques as were used to find $V_{NLL}$ in eq.~\eqref{eq64}.  However, since the three
loop RG functions have not been computed for the standard model, we will not pursue this calculation further.

\section*{Appendix 2: The Derivative Expansion of the Effective Action}

This paper has been concerned with contributions to the effective action coming from the first few terms in the derivative expansion when the
background field is either a scalar or vector field \cite{17}.  In this appendix we show how terms in this derivative expansion can be computed.
Operator regularization \cite{18} will be used in calculation. This technique has the advantages of not explicitly breaking any classical symmetries of
the theory (since no regulating parameter is inserted into the initial action) and of avoiding all explicit divergences at every stage of the calculation.

To illustrate this technique, we first consider a simple scalar model with a classical action
\begin{equation}
S^{(0)} = \int d^4x \left( - \frac{1}{2} (\partial_\mu \phi)^2 - \frac{1}{2} m^2\phi^2 - \frac{1}{6} \mu\phi^3 - \frac{1}{24} \lambda \phi^4\right)
\label{eqa1}
\end{equation}
If we split $\phi$ into the sum of a background part $f$ and a quantum fluctuation $h$ then performing the path integral over the quantum fluctuation leads to the
one loop contribution to the effective action
\begin{equation}
iS^{(1)} = -\frac{1}{2} tr\ln (p^2 + m^2 + \mu f + \frac{1}{2} \lambda f^2).\quad (p \equiv -i\partial)
\label{eqa2}
\end{equation}
Regulating the logarithm in eq.~\eqref{eqa2} using the zeta function \cite{18}
\begin{equation}
\left. \ln H = - \frac{d}{ds}\right|_0 H^{-s}
\left.  = - \frac{d}{ds}\right|_0 \frac{1}{\Gamma (s)} \int_0^\infty dit (it)^{s-1}e^{-iHt}
\label{eqa3}
\end{equation}
we see that eq.~\eqref{eqa2} can be written
\begin{equation}
\left. iS^{(1)} = \frac{1}{2} \frac{d}{ds}
\frac{1}{\Gamma (s)} \int_0^\infty dit (it)^{s-1} tr
\left\lbrace\exp -i (p^2 + m^2 + \mu f + \frac{1}{2} \lambda f^2)t\right\rbrace\right|_0~.
\label{eqa5}
\end{equation}
If now $f \rightarrow v + f$ where $v$ is a constant, and if $H = H_0 + H_1$ where
\begin{gather}
H_0 = p^2 + m^2 + \mu v + \frac{1}{2} \lambda v^2
\label{eqa6}
\\
H_1 = (\mu + \lambda v)f + \frac{\lambda f^2}{2}
\label{eqa7}
\end{gather}
then upon applying the Schwinger expansion \cite{18}
\begin{equation}
tr\, e^{-i(H_0+H_1)t} = tr\left[ e^{-iH_0t} + (-it)H_1 e^{-iH_0t}
+  \frac{1}{2} (-it)^2 \int_0^1 du H_1 e^{-i(1-u)H_0t} H_1 e^{-iuH_0t} + \ldots \right]
\label{eqa8}
\end{equation}
and keeping terms at most quadratic in $f$ we obtain
\begin{equation}
\begin{split}
iS^{(1)}_2 = \frac{1}{2} \frac{d}{ds}
\frac{\kappa^{2s}}{\Gamma(s)} \int_0^\infty dit (it)^{s-1}  tr & \Biggl\{  (-it)e^{-iH_0t} \left[
(\mu + \lambda v)f + \frac{\lambda f^2}{2}\right]\Biggr.
\\
&\left. \Biggl. + \frac{(-it)^2}{2} \int_0^1 du \;e^{-i(1-u)H_0t} (\mu + \lambda v)fe^{-iuH_0t} (\mu + \lambda v)f\Biggr\}
\right|_0
\end{split}
\label{eqa9}
\end{equation}
where $\kappa^{2s}$ is a dimensionful parameter inserted to ensure that $S^{(1)}$ is dimensionless (One could have introduced $\kappa^{2}$ in eq.~\eqref{eqa2} to keep the argument
of the logarithm dimensionless in that equation.).

The functional trace in eq.~\eqref{eqa9} can most easily be computed using momentum eigenstates $|p>$, $|q>$ and configuration eigenstates $|x>$, $|y>$ wherein $n$
dimensions $(2\pi)^{x/2} <x|p> = e^{ip\cdot x}$ so that
{\allowdisplaybreaks
\begin{gather}
\begin{split}
iS_2^{(1)} = \frac{1}{2} \frac{d}{ds} \frac{\kappa^{2s}}{\Gamma (s)} &\int_0^\infty dit(it)^{s-1}\, e^{-i(m^2+\mu v + \frac{1}{2} \lambda v^2)t}
\\
&\Biggl\{ (-it)\int dpdx  <p|e^{-ip^2t} |x><x|(\mu + \lambda v)f + \frac{\lambda f^2}{2} |p>  \Biggr.
\\
&+ \frac{1}{2} (-it)^2 \int dpdq dxdy \int_0^1 du <p|e^{-i(1-u)p^2t} |x><x|(\mu + \lambda v)f|q>
\\
&\qquad\left.\Biggl. <q|e^{-iuq^2} |y><y| (\mu + \lambda v)f|p>\Biggr\}\right|_0
\end{split}
\\
\begin{split}
iS_2^{(1)}
= \frac{1}{2} \frac{d}{ds} \frac{\kappa^2s}{\Gamma (s)} &\int_0^\infty dit\; e^{-i(m^2+\mu v + \frac{1}{2} \lambda v^2)t}
\Biggl\{
-(it)^{s} \int \frac{dpdx}{(2\pi)^4} e^{-ip^2t}
 \left[ (\mu + \lambda v)f(x) + \frac{\lambda}{2} f^2 (x)\right] \Biggr.
\\
& \left.\Biggl.+ \frac{1}{2} (it)^{s+1} (\mu + \lambda v)^2 \int \frac{dpdqdxdy}{(2\pi)^8}
e^{-i[(1-u)p^2+uq^2]t} e^{-i(p-q)\cdot (x-y)} f(x)f(y)\Biggr\}\right|_0 .
\end{split}
\label{eqa10}
\end{gather}
}
To obtain those terms which contribute to the effective action at one loop order which are second order in derivatives of the background field, we expand
$f(y)$ about $x$ up to second order so that
\begin{equation}
\begin{split}
\int &\frac{dpdqdxdy}{(2\pi)^8}  e^{-i[(1-u)p^2+uq^2]t} e^{-i(p-q)\cdot (x-y)} f(x)f(y)
\\
&\approx \int  \frac{dpdqdxdy}{(2\pi)^8} e^{-i[(1-u)p^2+uq^2]t} e^{-i(p-q)\cdot (x-y)}
 f(x)\left[f(x) + (x-y)^\alpha f_{,\alpha}(x) + \frac{1}{2}(x-y)^\alpha(x-y)^\beta f_{,\alpha\beta} (x)\right].
\end{split}
\label{eqa11}
\end{equation}
If now we write in eq.~\eqref{eqa11}
\begin{gather}
(x-y)^\alpha  e^{-i(p-q)\cdot(x-y)} = -i \frac{\partial}{\partial q^\alpha} e^{-i(p-q)\cdot (x-y)}
\label{eqa12}
\\
(x-y)^\alpha  (x-y)^\beta e^{-i(p-q)\cdot(x-y)} = (-i)^2 \frac{\partial}{\partial q^\alpha} \frac{\partial}{\partial q^\beta}
e^{-i(p-q)\cdot (x-y)}
\label{eqa13}  \end{gather}
and then perform an integration by parts with respect to $q$ we find that
\begin{equation}
\begin{split}
 iS_2^{(2)} = \frac{1}{2} \frac{d}{ds} \frac{\kappa^2s}{\Gamma (s)} &\int_0^\infty dit\; e^{-i(m^2+\mu v + \frac{1}{2} \lambda v^2)t}
\Biggl\{ -(it)^s \frac{i}{(4\pi it)^s}
\int dx \left[(\mu + \lambda v)f(x) + \frac{1}{2} \lambda f^2(x)\right]\Biggr.
\\
&\left.\Biggl. + \frac{1}{2} (it)^{s+1} \int_0^1 du(\mu + \lambda v)^2
\left[ f^2(x) + (it) u(1-u) f(x) \partial^2 f(x)\right]\Biggr\} \right|_0
\end{split}
\label{eqa14}
\end{equation}
where we have used the integral
\begin{equation}
\int \frac{d^np}{(2\pi)^n} e^{-ip^2t} = \frac{i}{(4\pi it)^{n/2}} .
\label{eqa15a}
\end{equation}

The integrals over $t$ and $u$ are now standard and we end up with
\begin{equation}
\begin{split}
iS_2^{(1)} = \frac{i}{32\pi^2} &\int dx \Biggl\{ \left[(\mu + \lambda v)f(x) + \frac{1}{2} \lambda f^2(x)\right]\left[m^2 + \lambda v
+ \frac{1}{2} \lambda v^2\right]  \left[ 1 - \ln\left(\frac{m^2+\mu v + \frac{1}{2} \lambda v^2}{\kappa^2}\right)\right]\Biggr.
\\
&\Biggl. - \frac{1}{2} (\mu + \lambda v)^2 f^2 (x) \ln
\left(\frac{m^2 + \mu v+\frac{1}{2} \lambda v^2}{\kappa^2}\right)
+ \frac{1}{2} \frac{(\mu + \lambda v)^2 f(x)\partial^2f(x)}{(m^2 + \mu v+\frac{1}{2} \lambda v^2)}\Biggr\} .
\end{split}
\label{eqa15}
\end{equation}
Eq.~\eqref{eqa15} agrees with what was obtained using different techniques in ref.~\cite{17}.

The approach outlined for the simple scalar model of eq.~\eqref{eqa1} can easily be applied to compute terms in the derivative expansion of the
effective action in more complicated models.  For scalar electrodynamics with the classical action
\begin{equation}
S_\phi = \int d^4x \left[ - (\partial_\mu + ieV_\mu)\phi^* (\partial^\mu - ieV^\mu)\phi - \lambda (\phi^*\phi)^2
- \frac{1}{4} (\partial_\mu V_\nu -
\partial_\nu V_\mu)^2\right]
\label{eqa16}
\end{equation}
we again let $\phi = f + h$ where $f$ is the background field.  Using the gauge fixing term
\begin{equation}
S_{gf} = -\frac{1}{2\alpha} \int d^4x \left[\partial \cdot V + \frac{ie\alpha}{2} (f^*h - fh^*)\right]^2
\label{eqa17}
\end{equation}
and the attendant ghost action
\begin{equation}
S_{gh} = \int d^4x \;\overline{c} \left[\partial^2 - \frac{1}{2} e^2\alpha (2f^*f  + f^*h + fh^*)\right]
\label{eqa18}
\end{equation}
we find that the one loop effective action is given by
\begin{equation}
\begin{split}
iS^{(1)} =& \ln\det\left[ p^2 + e^2\alpha (f_1^2 + f_2^2)\right]
\\
&- \frac{1}{2} \ln\det\left[
\begin{array}{ccc}
p^2 + 3\lambda f_1^2 + (\lambda + \alpha e^2)f_2^2 & (2\lambda - \alpha e^2)f_1f_2 & -2ef_{2,\,\nu}\\
(2\lambda - \alpha e^2)f_1f_2 & p^2 +(\lambda + \alpha e^2)f_1^2 + 3\lambda f_2^2   & 2ef_{1,\,\nu}\\
-2ef_{2,\mu} & 2ef_{1,\mu} & p^2(T + \frac{1}{\alpha} L)_{\mu\nu} + e^2(f_1^2 + f_2^2)g_{\mu\nu}
\end{array}
\right]
\label{eqa19}
\end{split}
\end{equation}
where $f_1$ and $f_2$ are the real and imaginary parts of $f$ and $T_{\mu\nu} = g_{\mu\nu} - p_\mu p_\nu/p^2$, $L_{\mu\nu} = p_\mu p_\nu/p^2$ are a complete
set of orthogonal projection operators.

Operator regularization can now be applied to this expression in the same way as it was applied to eq.~\eqref{eqa2}; after the replacement $f_1 \rightarrow v + f_1$
the Schwinger expansion is used to obtain all terms second order in $f_1$ and $f_2$ and these fields can then be expanded out to second order in a Taylor expansion
about some point $x$.  One could also expand in powers of the external field strength and its derivatives.

\end{document}